\documentclass[graybox]{svmult}

\usepackage{type1cm}        
\usepackage{makeidx}         
\usepackage{graphicx}        
\usepackage{multicol}        
\usepackage[bottom]{footmisc}

\usepackage{newtxtext}       %
\usepackage[varvw]{newtxmath}       
\usepackage{multirow}
\usepackage{adjustbox}
\usepackage{orcidlink}

\makeindex 

\begin{document}


\graphicspath{{Figures/}}
\title*{Confronting primordial black holes with LIGO-Virgo-KAGRA and the Einstein Telescope}

\author{Zu-Cheng Chen\orcidlink{0000-0001-7016-9934} and Alex Hall\orcidlink{0000-0002-3139-8651}}
\institute{Zu-Cheng Chen \at Department of Physics, Hunan Normal University, China, \email{zuchengchen@hunnu.edu.cn} \and Alex Hall \at Institute for Astronomy, University of Edinburgh, United Kingdom, \email{ahall@roe.ac.uk}}

\maketitle\label{chapter:LVK-ET}

\abstract*{The detection of gravitational waves (GWs) from binary black hole (BBH) coalescences by the LIGO-Virgo-KAGRA (LVK) Collaboration has raised fundamental questions about the genesis of these events. In this chapter, we explore the possibility that PBHs, proposed candidates for dark matter, may serve as the progenitors of the BBHs observed by LVK. 
Employing a Bayesian analysis, we constrain the PBH model using the LVK third GW Transient Catalog (GWTC-3), revealing that stellar-mass PBHs cannot dominate cold dark matter.
Considering a mixed population of astrophysical black holes (ABHs) and PBHs, we determine that approximately $1/4$ of the detectable events in the GWTC-3 can be attributed to PBH binaries. 
We also forecast detectable event rate distributions for PBH and ABH binaries by the third-generation ground-based GW detectors, such as the Einstein Telescope, offering a potential avenue to distinguish PBHs from ABHs based on their distinct redshift evolutions.
}

\abstract{The detection of gravitational waves (GWs) from binary black hole (BBH) coalescences by the LIGO-Virgo-KAGRA (LVK) Collaboration has raised fundamental questions about the genesis of these events. In this chapter, we explore the possibility that PBHs, proposed candidates for dark matter, may serve as the progenitors of the BBHs observed by LVK. 
Employing a Bayesian analysis, we constrain the PBH model using the LVK third GW Transient Catalog (GWTC-3), revealing that stellar-mass PBHs cannot dominate cold dark matter.
Considering a mixed population of astrophysical black holes (ABHs) and PBHs, we determine that approximately $1/4$ of the detectable events in the GWTC-3 can be attributed to PBH binaries. 
We also forecast detectable event rate distributions for PBH and ABH binaries by the third-generation ground-based GW detectors, such as the Einstein Telescope, offering a potential avenue to distinguish PBHs from ABHs based on their distinct redshift evolutions.
}

\section{Introduction}
\label{chap25:sec:intro}
In recent years, the detection of gravitational waves (GWs) from compact binary coalescences by the LIGO-Virgo-KAGRA (LVK) Collaboration has ushered in a new era of GW astronomy~\cite{LIGOScientific:2018mvr,LIGOScientific:2020ibl,LIGOScientific:2021djp}. The latest release of the third GW Transient Catalog (GWTC-3) by LVK reports approximately $90$ GW events detected during their first three observing runs, with a majority identified as binary black hole (BBH) mergers exhibiting a broad mass distribution. The heaviest event, GW190521~\cite{LIGOScientific:2020iuh}, involves component masses $m_1 = 85^{+21}_{-14}M_\odot$ and $m_2 = 66^{+17}_{-18}M_\odot$, residing within the upper black hole mass gap believed to originate from pulsation pair-instability supernovae~\cite{Marchant:2018kun}. The lower cutoff of this mass gap is currently estimated to be around $50\pm 4M_\odot$~\cite{Belczynski:2016jno,Marchant:2018kun,Farmer:2019jed,Farmer:2020xne,Marchant:2020haw}. Even after accounting for measurement uncertainties, this observation strongly suggests that the primary mass of GW190521 lies within the mass gap, challenging the idea of a direct stellar progenitor origin~\cite{Anagnostou:2020tta} within the conventional stellar evolution scenario for astrophysical black holes (ABHs).

As an alternative explanation for the observed BBHs, primordial black holes (PBHs) have gained attention alongside conventional ABHs. Formed in the early Universe through the gravitational collapse of primordial density fluctuations~\cite{Hawking:1971ei,Carr:1974nx} (see also Chapter~4), PBHs serve as a plausible explanation for LVK BBHs~\cite{Bird:2016dcv,Sasaki:2016jop} and are considered candidates for cold dark matter (CDM)~\cite{Carr:2016drx} and seeds for galaxy formation~\cite{Bean:2002kx,Kawasaki:2012kn}. 
Moreover, the recent pulsar timing observation~\cite{Barr:2024wwl} of the eccentric binary millisecond pulsar, PSR J0514$-$4002E, suggests that its companion could potentially be a PBH~\cite{Chen:2024joj}.
The formation of PBHs is inevitably associated with the generation of GWs induced by primordial scalar perturbations~\cite{Saito:2008jc,Cai:2018dig,Yuan:2019udt,Yuan:2019wwo,Yuan:2019fwv,Chen:2019xse,DeLuca:2019ufz,Bartolo:2018rku,Bartolo:2018evs} (see also Chapter~18). These induced GWs may be detectable with pulsar timing array experiments~\cite{NANOGrav:2023gor,EPTA:2023fyk,Xu:2023wog,Reardon:2023gzh,Chen:2019xse,Liu:2023ymk,Jin:2023wri,Liu:2023pau,Liu:2023hpw,Chen:2024fir,Yi:2023tdk,You:2023rmn}.

In this chapter, we explore the plausibility of PBHs as an explanation for the BBH events observed by LVK (see Ref.~\cite{LISACosmologyWorkingGroup:2023njw} for a recent review).
The observed rate of GW events puts a strong constraint on models in which a significant proportion of CDM is in PBHs. Despite this, a subdominant population of PBHs is permitted by the data, and such a population may be able to account for features in the source population that are not easily explained by ABHs.
Our analysis indicates that the abundance of stellar-mass PBHs in CDM, $f_\mathrm{PBH}$, cannot exceed $\mathcal{O}(10^{-3})$. 
Furthermore, the GW data suggest that not all BBH events detected so far can be solely explained by a single formation channel, whether astrophysical~\cite{Zevin:2020gbd} or primordial~\cite{Hall:2020daa,Wong:2020yig}. We, therefore, also present a comprehensive Bayesian inference study of the GWTC-3 catalog, incorporating a mixed ABH + PBH model.

We also explore the potential for distinguishing between ABHs and  PBHs using upcoming third-generation ground-based GW detectors, such as the Einstein Telescope (ET)~\cite{Punturo:2010zz}. The population properties of PBH binaries differ significantly from those of ABH binaries. Notably, the merger rate of PBH binaries increases with redshift ($z$), while the merger rate of ABH binaries, which follows the star formation rate, peaks around $z\sim2$ and then rapidly decreases. This characteristic feature can serve as a distinguishing factor between PBHs and ABHs~\cite{Chen:2019irf,Mukherjee:2021ags}.

\section{PBH models and Bayesian methodology}
\label{chap25:sec:howto}

In this section, we begin with an overview of the calculation of the merger rate density of PBH binaries by considering the four most commonly used PBH mass functions in the literature. Then, we present the Bayesian inference framework employed in data analyses to constrain the various PBH models.

\subsection{Merger rate density distribution of PBH binaries}
\label{chap25:subsec:mergerrate}

In this subsection, we outline the calculation of the PBH merger rate density. Throughout this chapter, our focus is solely on the scenario that PBH binaries are formed in the early Universe. For the late Universe scenario and a comprehensive discussion of the merger rate density of PBH binaries, readers are directed to Chapter~17. 

The sample of BBHs observed by LVK implies a broad distribution of BH masses, prompting the consideration of an extended mass function for PBHs. We impose the normalization condition on the probability distribution function of PBH mass, $P(m)$, ensuring that
\begin{equation}
\label{chap25:eq:norm}
\int_{0}^{\infty} P(m)\, \mathrm{d} m = 1.
\end{equation} 
Assuming that the fraction of CDM in PBHs is $f_\mathrm{PBH} \equiv \Omega_\mathrm{PBH} / \Omega_\mathrm{CDM}$, then the fraction of PBHs in the non-relativistic matter within the mass interval $(m, m+\mathrm{d} m)$ is~\cite{Chen:2018rzo}
\begin{equation} 
f_\mathrm{CDM} f_\mathrm{PBH}\, P(m)\, \mathrm{d} m,
\end{equation} 
where the coefficient $f_\mathrm{CDM}\approx 0.85$ accounts for the fraction of CDM in the non-relativistic matter, encompassing both CDM and baryons. 

We can now proceed to estimate the merger rate density of PBH binaries. We assume that PBHs are initially randomly distributed, following a spatial Poisson distribution in the early Universe, a condition that holds when they decouple from the cosmic background evolution~\cite{Nakamura:1997sm,Sasaki:2016jop,Ali-Haimoud:2017rtz}. Pairs of PBHs acquire angular momentum due to the gravitational torque exerted by other PBHs, ultimately culminating in the formation of a PBH binary upon decoupling from cosmic expansion.
Subsequently, the binary undergoes the emission of gravitational radiation, ultimately resulting in a merger GW event potentially detectable by GW detectors.
The merger rate density of a PBH binary with masses $m_1$ and $m_2$ at cosmic time $t$, $\mathcal{R}(t, m_1, m_2)$, is expressed as~\cite{Chen:2018czv,Raidal:2018bbj,Liu:2018ess}
\begin{equation}
\label{chap25:eq:cR}
\mathcal{R}(t, m_1, m_2) = \frac{1.6\times 10^{-6}}{\mathrm{Gpc^3\,yr}} \left(\frac{t}{t_0}\right)^{-\frac{34}{37}} f_\mathrm{PBH}^\frac{53}{37} \eta^{-\frac{34}{37}} \left(\frac{M}{M_\odot}\right)^{-\frac{32}{37}} P(m_1) P(m_2),
\end{equation}
where $t_0$ is the present cosmic time, $M=m_1 + m_2$ is the total mass, and $\eta\equiv m_1 m_2/(m_1+m_2)^2$.
The contribution of hierarchical mergers~\cite{Liu:2019rnx} has been neglected, given its subdominance as constrained by GWTC~\cite{Wu:2020drm,Liu:2022iuf}.
Additionally, the total merger rate can be obtained by integrating the component masses as
\begin{equation}
\label{chap25:eq:Rt}
R(t) = \int \mathcal{R}(t, m_1, m_2)\, \mathrm{d} m_1\, \mathrm{d} m_2.
\end{equation} 
We will frequently switch between cosmic time $t$ and redshift $z$.
The cosmic time $t$ and redshift $z$ are related by
\begin{equation}
t(z) = \int_{z}^{\infty} \frac{dz'}{H(z') (1+z')},
\end{equation}
where $H(z)$ is the Hubble parameter. We neglect contributions from radiation and neutrinos, given the limited sensitivity of current ground-based GW detectors to a small redshift range. When converting between luminosity distances, times, and redshifts, we adopt the best-fit cosmological model of Planck 2018~\cite{Planck:2018vyg}.

\begin{figure}[tbp!]
\centering
\includegraphics[width=\textwidth]{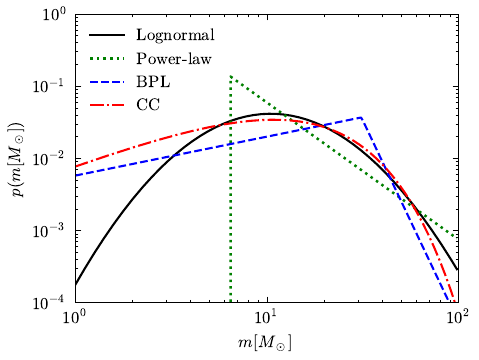}
\caption{An illustrative depiction presenting the four PBH mass distributions explored in this chapter: lognormal, power-law, broken power-law (BPL), and critical collapse (CC) functional forms. All four PBH mass functions have been normalized to unity, satisfying Eq.~\eqref{chap25:eq:norm}.}
\label{chap25:fig:mass_funcs}
\end{figure}

To constrain the PBH scenario, one must place some a priori constraints on the form of the PBH mass function, $P(m)$. The form of the mass function is sensitive to the details of PBH formation. Throughout this chapter we consider four PBH mass functions frequently employed in the literature: the lognormal, power-law (with a lower-mass cutoff), broken power-law, and critical collapse distributions. In Fig.~\ref{chap25:fig:mass_funcs}, we provide a visual illustration of these four PBH mass distributions.
In the rest of the subsection, we will introduce these mass functions and briefly describe their key properties.

\subsubsection{Lognormal mass function}
\label{cha25:susubsec:log}

A lognormal PBH mass function is defined by~\cite{Dolgov:1992pu}
\begin{equation}
\label{chap25:eq:log}
P(m) = \frac{1}{\sqrt{2\pi} \sigma m} \exp \left(-\frac{\ln^2\left(m/M_c\right)}{2\sigma^2}\right),
\end{equation}
where $M_c$ is the median mass and $\sigma$ characterizes the width of the mass distribution.
The lognormal mass function proves to be a versatile approximation for a wide range of extended mass distributions, especially when PBHs originate from a smooth, symmetric peak in the inflationary power spectrum under the slow-roll approximation~\cite{Green:2016xgy,Carr:2017jsz,Kannike:2017bxn}.
A PBH population with a lognormal mass function therefore requires at least three parameters to model the binary merger rate, $\Lambda = \{M_c, \sigma, f_\mathrm{PBH}\}$.

\subsubsection{Power-law mass function}
\label{cha25:susubsec:power}

The next model we consider is a PBH mass function characterized by the power-law form~\cite{Carr:1975qj}
\begin{equation}
\label{chap25:eq:power} 
P(m) = \frac{\alpha-1}{M_\mathrm{min}} \left(\frac{m}{M_\mathrm{min}}\right)^{-\alpha},
\end{equation}
where $M_\mathrm{min}$ denotes a lower-mass cutoff and $\alpha$ is the power-law index.
We require $\alpha>1$ such that $\int_{M_\mathrm{min}}^{\infty} P(m) \mathrm{d} m$ will not be divergent.
The power-law mass function typically emerges from a broad or flat power spectrum of curvature perturbations~\cite{DeLuca:2020ioi} during the radiation-dominated era \cite{Carr:2016drx,Carr:2017jsz}.
In this case, three hyperparameters, $\Lambda = \{M_\mathrm{min}, \alpha, f_\mathrm{PBH}\}$, need to be determined through the analysis of GW data. 

\subsubsection{Broken power-law mass function}
\label{cha25:susubsec:bpl}

A refinement of the power-law model is to consider a broken power-law form~\cite{Deng:2021ezy} given by
\begin{equation}
	P(m)= \left(\frac{m_*}{\alpha_1+1} + \frac{m_*}{\alpha_2-1}\right)^{-1} \begin{cases} (\frac{m}{m_*})^{\alpha_1}, & m<m_* \\ (\frac{m}{m_*})^{-\alpha_2}, & m>m_*\end{cases},
\end{equation}
where $m_*$ is the peak mass of $m P(m)$. Here, $\alpha_1>0$ and $\alpha_2>1$ are two power-law indices. The broken power-law mass function is a generalization of the power-law form and can be realized if PBHs are formed by vacuum bubbles nucleating during inflation via quantum tunneling~\cite{Deng:2021ezy}.
In this case, four hyperparameters, $\Lambda = \{m_*, \alpha_1, \alpha_2, f_\mathrm{PBH}\}$, need to be determined through the analysis of GW data.

\subsubsection{Critical collapse mass function}
\label{cha25:susubsec:cc}

Lastly, we consider a PBH mass function with the critical collapse form~\cite{Niemeyer:1997mt,Yokoyama:1998xd,Carr:2016hva,Gow:2020cou}
\begin{equation}
P(m)=\frac{\alpha^2\,  m^\alpha}{M_\mathrm{f}^{1+\alpha}\, \Gamma(1 / \alpha)} \exp \left(-(m/M_\mathrm{f})^{\alpha}\right),
\end{equation}
where $\alpha$ is a universal exponent related to the critical collapse of radiation, and $M_{\mathrm{f}}$ is a mass scale approximately of the order of the horizon mass at the collapse epoch~\cite{Carr:2016hva}.
This mass spectrum does not exhibit a lower mass cutoff; however, it experiences exponential suppression beyond the mass scale of $M_\mathrm{f}$. The critical collapse mass function is closely related to a power spectrum of density fluctuations resembling a $\delta$-function~\cite{Niemeyer:1997mt,Yokoyama:1998xd,Carr:2016hva,Gow:2020cou}.
In this case, three hyperparameters, $\Lambda = \{M_\mathrm{f}, \alpha, f_\mathrm{PBH}\}$, need to be determined through the analysis of GW data.

\subsection{Bayesian formalism}
\label{chap25:subsec:bayes}

In the PBH context, population studies using the Bayesian formalism have become standard practice~\cite{Chen:2018rzo,Hall:2020daa,Hutsi:2020sol,DeLuca:2021wjr,Franciolini:2021tla,Chen:2021nxo,Chen:2022fda,Liu:2022iuf,Zheng:2022wqo}. 
This subsection gives an overview of the Bayesian formalism employed in the analysis of GW data to infer model parameters. We refer to \cite{Thrane:2018qnx} for a pedagogical introduction to Bayesian inference in GW astronomy. With the recent accumulation of almost $100$ GW observations from the LVK Collaboration~\cite{KAGRA:2021vkt}, the study of GW sources is swiftly transitioning to the domain of ``population inference"~\cite{LIGOScientific:2018jsj,Chen:2018rzo,Chen:2019irf,LIGOScientific:2020kqk,Chen:2021nxo,KAGRA:2021duu,Chen:2022fda,Liu:2022iuf,Zheng:2022wqo,You:2023ouk}. This involves addressing selection biases introduced by the detectors' sensitivity, particularly concerning GW frequency and binary parameters like masses and redshift. Beyond correcting for selection bias, GW population inference faces the challenge of dealing with uncertain source parameter measurements due to noise in the heterogeneous data. This requires a specialized statistical framework to reconstruct population properties, with current techniques often leveraging hierarchical Bayesian inference~\cite{Mandel:2018mve,Vitale:2020aaz}. To exemplify how the machinery of BBH population analysis works in detail, in the remainder of this chapter we will employ a hierarchical Bayesian approach to deduce BBH population parameters from GW data, accounting for the uncertainty in estimating individual event parameters. 

The merger rate density, as expressed in Eq.~\eqref{chap25:eq:cR}, corresponds to measurements made in the BBH source frame of reference. The source frame is the coordinate system attached to the astrophysical source of GW, while the detector frame is the coordinate system associated with the GW detector. In other words, the source frame is linked to the astrophysical properties of the GW source, while the detector frame is associated with the measurements recorded by the GW detector. To facilitate analysis, a conversion into the detector frame is required using the transformation given by
\begin{equation}
\label{chap25:eq:Rpop}
\mathcal{R}_{\mathrm{pop}}(\theta|\Lambda) = \frac{1}{1+z} \frac{dV_\mathrm{c}}{dz} \mathcal{R}(\theta|\Lambda),
\end{equation}
where $z$ is the cosmological redshift, and $\theta\equiv \{z, m_1, m_2\}$ constitutes the parameter set defining individual BBH events.
In this chapter, we focus on the redshift and mass distributions and ignore the spin distribution.
The collection of parameters $\Lambda$ encompasses $f_\mathrm{PBH}$ and parameters derived from the mass function $P(m)$. Additionally, $dV_\mathrm{c}/dz$ represents the differential comoving volume. The incorporation of the factor $1/(1 + z)$ serves to convert time increments from the source frame to the detector frame.

Given the observed data $\textbf{d} = \{d_1, d_2, \cdots, d_{N_{\mathrm{obs}}}\}$, comprising $N_{\mathrm{obs}}$ BBH merger events, we formulate the total number of events as an inhomogeneous Poisson process, leading to the likelihood \cite{Loredo:2004nn,Thrane:2018qnx,Mandel:2018mve}
\begin{equation}
\label{chap25:eq:L1}
	\mathcal{L}(\textbf{d}|\Lambda) \propto N_{\mathrm{exp}}^{N_{\mathrm{obs}}} e^{-N_{\mathrm{exp}}} \prod_{i=1}^{N_{\mathrm{obs}}} \frac{\int \mathcal{L} (d_{i}| \theta)\, \mathcal{R}_{\mathrm{pop}}(\theta|\Lambda) d \theta}{\xi(\Lambda)},
\end{equation}
where $N_{\exp}\equiv N_{\exp}(\Lambda)$ represents the expected number of detections over the observation timespan. Here, $\mathcal{L} (d_{i}|\theta)$ denotes the likelihood for the individual $i$th GW event, derived from the posterior of the individual event by reweighing with the prior on $\theta$. Furthermore, $\xi(\Lambda)$ quantifies selection biases for a population with parameters $\Lambda$ and is defined as
\begin{equation}
\label{chap25:eq:xi}
\xi(\Lambda) = \int P_{\mathrm{det}}(\theta)\, \mathcal{R}_{\mathrm{pop}}(\theta|\Lambda)\, \mathrm{d} \theta,
\end{equation}
where $0 < P_{\mathrm{det}}(\theta) < 1$ denotes the detection probability~\cite{OShaughnessy:2009szr}, a function of the source parameters $\theta$.

In practice, the estimation of $\xi(\Lambda)$ is achieved using simulated injections~\cite{ligo_scientific_collaboration_and_virgo_2021_5546676}. Then, 
$\xi(\Lambda)$ in Eq.~\eqref{chap25:eq:xi} is approximated through a Monte Carlo integral over found injections~\cite{KAGRA:2021duu} as
\begin{equation}
	\xi(\Lambda) \approx \frac{1}{N_{\mathrm{inj}}} \sum_{j=1}^{N_{\text {found }}} \frac{\mathcal{R}_{\mathrm{pop}}(\theta_{j} | \Lambda)}{p_{\mathrm{draw}}(\theta_j)},
\end{equation}
where $N_{\text{inj}}$ stands for the total number of injections, $N_{\text{found}}$ is the count of successfully detected injections, and $p_{\mathrm{draw}}$ represents the probability density function from which the injections are drawn. Utilizing the posterior samples from each event, we compute the hyper-likelihood~\eqref{chap25:eq:L1} as
\begin{equation}
\label{chap25:eq:L2}
\mathcal{L}(\textbf{d}|\Lambda) \propto N_{\mathrm{exp}}^{N_{\mathrm{obs}}} e^{-N_{\mathrm{exp}}} \prod_{i=1}^{N_{\mathrm{obs}}} \frac{1}{\xi(\Lambda)} \left\langle \frac{\mathcal{R}_{\mathrm{pop}}(\theta|\Lambda)}{\pi(\theta)} \right\rangle,
\end{equation}
where $\langle\cdots\rangle$ denotes the weighted average over posterior samples of $\theta$. The denominator $\pi(\theta)$ refers to the priors on source parameters used to construct the posterior of individual events.

According to Bayes's theorem, the posterior distribution for the hyperparameter $\Lambda$ is expressed as
\begin{equation}
    p(\Lambda | \textbf{d})=\frac{\mathcal{L}(\textbf{d} | \Lambda) \pi(\Lambda)}{\mathcal{Z}},
\end{equation}
$\pi(\Lambda)$ denotes the prior distribution for $\Lambda$, and $\mathcal{Z}$ is a normalization factor  referred to as the ``evidence"
\begin{equation}
\mathcal{Z} \equiv \int \mathrm{d} \Lambda\, \mathcal{L}(\textbf{d} | \Lambda) \pi(\Lambda).
\end{equation}
In this chapter, we integrate the PBH population distribution~\eqref{chap25:eq:cR} into the \texttt{ICAROGW}~\cite{Mastrogiovanni:2021wsd} package to calculate the likelihood function~\eqref{chap25:eq:L2}. We utilize the \texttt{dynesty}~\cite{Speagle:2019ivv} sampler, accessed through \texttt{Bilby}~\cite{Ashton:2018jfp,Romero-Shaw:2020owr}, to perform sampling across the parameter space.

The evidence, $\mathcal{Z}$, defined above plays a crucial role in model selection, addressing the question of which model is statistically favored by the data and to what extent. The Bayes factor, quantifying the ratio of evidence between two exclusive models, $\mathcal{M}_2$ and $\mathcal{M}_1$, serves as a quantifiable measure for model selection scores. It is expressed as
\begin{equation}
\mathrm{BF}^2_1 \equiv \frac{\mathcal{Z}_2}{\mathcal{Z}_1},
\end{equation}
with $\mathcal{Z}_2$ and $\mathcal{Z}_1$ denoting the evidence for $\mathcal{M}_2$ and $\mathcal{M}_1$, respectively. The evidence, being the average of the likelihood over the prior volume, naturally incorporates Occam’s razor principle, which states that, all else being equal, a model with fewer parameters is preferred. A substantial value of $\mathrm{BF}^2_1$ indicates a strong preference for model $\mathcal{M}_2$ over $\mathcal{M}_1$. Table~\ref{chap25:tab:BF} presents an interpretation of the Bayes factor in the model comparison~\cite{BF}.

\begin{table}[tbp!]
\centering
\begin{tabular}{ccl}
\hline
\hspace{3mm}BF\hspace{3mm} & \hspace{2mm}$\ln \mathrm{BF}$\hspace{3mm} & Strength of evidence \\
\hline$<1$ & $<0$ & Negative \\
$1-3$ & $0-1$ & Not worth more than a bare mention \\
$3-20$ & $1-3$ & Positive \\
$20-150$ & $3-5$ & Strong \\
$>150$ & $>5$ & Very strong \\
\hline
\end{tabular}
\caption{An interpretation of the Bayes factor in quantifying the model selection scores as given by Ref.~\cite{BF}.}
\label{chap25:tab:BF}
\end{table}

\section{Constraints from the GWTC-3 catalog}
\label{chap25:sec:gwtc3}

In this section, we first present the constraints on the population parameters using GWTC-3 for the individual scenarios of purely ABH population, purely PBH population, and a combined model containing both ABH and PBH populations. 

\begin{figure}[htb!]
\centering
\includegraphics[width=0.74\textwidth]{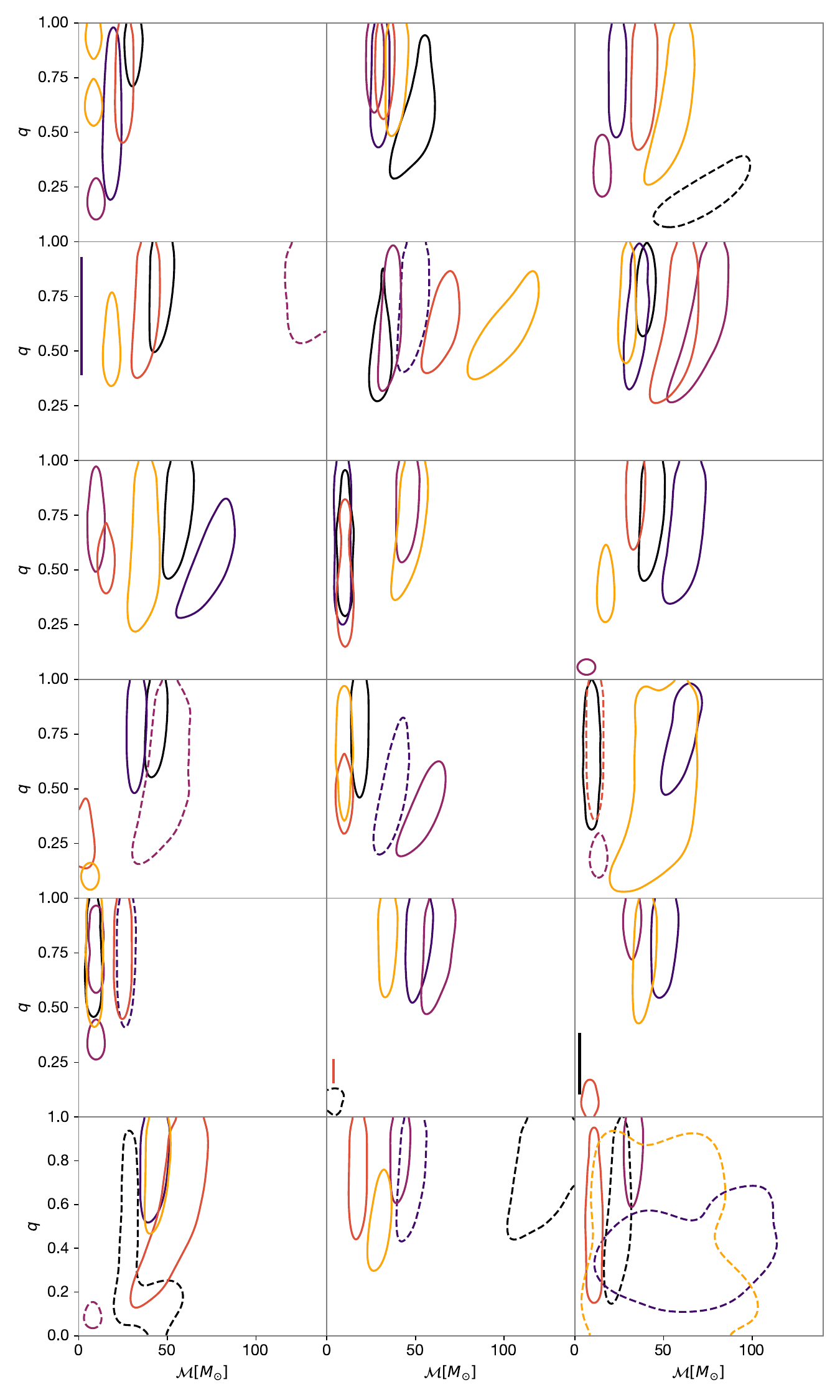}
\caption{Source parameter constraints for the $90$ CBCs that constitute GWTC-2.1 and GWTC-3. Each contour encloses $68\%$ of the posterior probability, and we show constraints projected into the plane of mass ratio $q$ and detector-frame chirp mass $\mathcal{M}$. Sources have been placed into groups of five according to their detection time, with the earliest detections (i.e.~those detected in O1) in the top-left panel and latest in the bottom-right (i.e.~those detected in O3b). Hotter, more luminant colours denote sources with a later detection date within its group of five. Sources with dashed contours have a false alarm rate (FAR) larger than 1 per year, namely $\mathrm{FAR} > 1\, \mathrm{yr}^{-1}$, and are hence excluded from the population analysis of~\cite{KAGRA:2021duu}. We have also excluded the binary neutron star event GW170817, but included GW200105\_162426, which was excluded from GWTC-3 due to its low $p_{\mathrm{astro}}$ but included in the populations analysis of Ref.~\cite{KAGRA:2021duu}.}
\label{chap25:fig:GWTC}       
\end{figure}

\begin{figure}[tbh!]
\centering
\includegraphics[width=\textwidth]{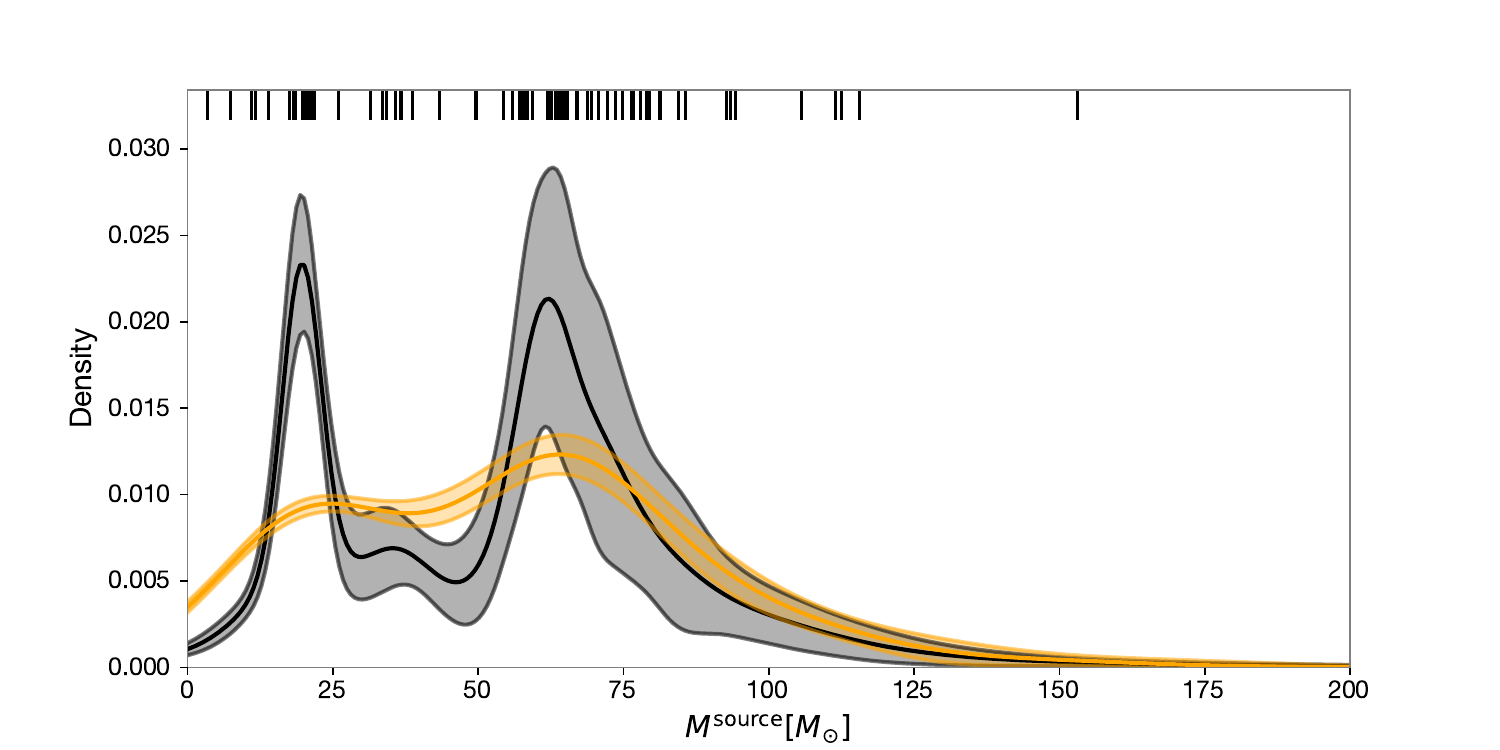}
\includegraphics[width=\textwidth]{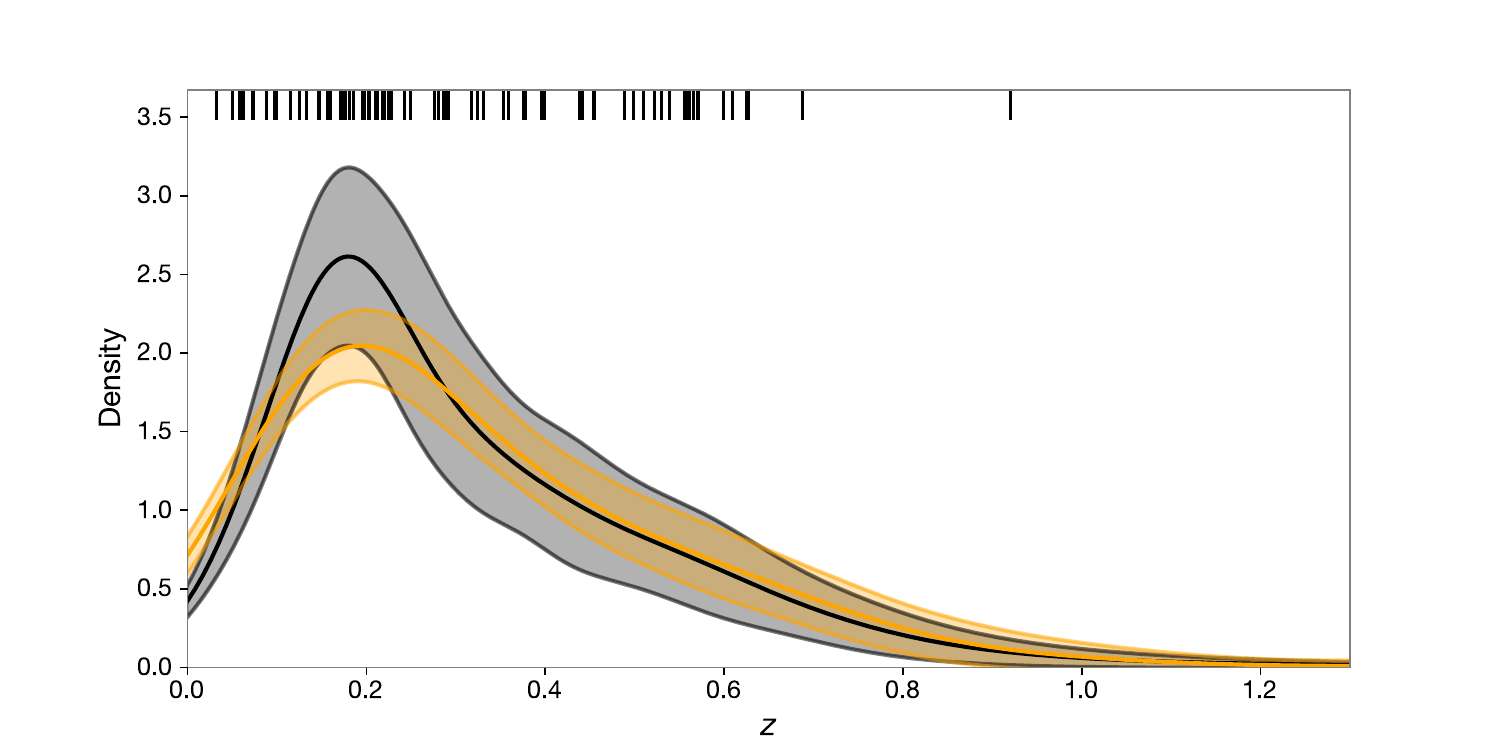}
\caption{Distribution of the source-frame total mass (top panel) and redshift (bottom panel) of CBCs in the GWTC catalog, excluding GW170817 and events with $\mathrm{FAR} > 1\, \mathrm{yr}^{-1}$. Posterior medians are shown as tick marks at the top of the figure. The black curve shows an adaptive Gaussian KDE constructed with~\texttt{awkde}\protect\footnotemark and using the algorithm of Ref.~\cite{Sadiq:2021fin}, the orange curve shows a standard Gaussian KDE with a global bandwidth set using Scott's rule. Both curves are mean KDEs over bootstrap resamples from the source posteriors, shaded regions are $2\sigma$ confidence intervals constructed from these resamples.}
\label{chap25:fig:mtot_kdes}       
\end{figure}

\footnotetext{\url{https://github.com/mennthor/awkde}}

As described above, the GWTC provides samples from the posterior distribution of parameters describing each GW event. These include both \emph{intrinsic} parameters, such as source-frame component masses and spins, and \emph{extrinsic} parameters such as matched filter signal-to-noise (SNR) and merger times. Bayesian inference of population models can then be performed by approximating the integral over source parameters $\theta$ in the likelihood function Eq.~\eqref{chap25:eq:L1} with the Monte Carlo estimator Eq.~\eqref{chap25:eq:L2}.

In the subsection, we will describe constraints from GWTC-3~\cite{KAGRA:2021vkt} (incorporating its previous iterations, see Refs.~\cite{LIGOScientific:2018mvr,LIGOScientific:2020ibl}) assuming the observed BBH merger events are drawn from a population of solely ABHs, solely PBHs, and finally a mixed ABH+PBH population.

\subsection{Constraints on the ABH population}
\label{chap25:subsubsec:ABH}

Population inference has been performed extensively with the GWTC3 data and its predecessors in the ABH context~\cite{LIGOScientific:2018mvr,Roulet:2018jbe, Roulet:2020wyq,LIGOScientific:2020kqk,Tiwari:2021yvr,KAGRA:2021duu}. Detailed modelling of the ABH merging binary population is challenging due to the plurality of possible formation channels and astrophysical uncertainties (for recent reviews, see Refs.~\cite{Mandel:2018hfr,Mandel:2021smh}). Partly for this reason, ABH population studies have primarily taken an agnostic approach to model inference, for example, by considering simple empirical parametric merger rate models. 

The best fitting empirical parametric models are those that can accommodate the primary features visible in the distribution of detector-frame chirp mass, as this is typical a well constrained parameter for a BBH merger event. The empirical distribution is qualitatively similar to that of the source-frame total mass, shown in the top panel of Figure~\ref{chap25:fig:mtot_kdes}; a peak in the merger rate for total masses around $20 \, M_\odot$, relatively few sources in the range $30$ -- $50 \, M_\odot$, a broad peak between $50 \, M_\odot$ and $80 \, M_\odot$, and a gradual suppression in the rate of sources with higher mass. While models with more parameters may be able to provide superior fits to the data, overfitting incurs penalties for such models when compared against simpler models in Bayesian model selection.

While detailed predictions of the merger rate density as a function of source masses and redshift are challenging in the ABH context, stellar physics makes several predictions for the black hole population of varying robustness to physical assumptions. The most secure of these is the lower limit on the mass of stellar black holes, thought to be equivalent to the maximum neutron star mass, around $2 \, M_\odot$~\cite{Tolman:1939jz,Oppenheimer:1939ne,Rezzolla:2017aly}. For black holes just heavier than this, a mass gap spanning the range $2$--$5\, M_\odot$ can develop~\cite{Belczynski:2011bn}, although this is subject to uncertainty based on the timescale for instability growth and supernova launch~\cite{Fryer:2011cx}. The discovery of the event GW190814, with a secondary component mass of $2.6\, M_\odot$~\cite{LIGOScientific:2020zkf} directly challenges the existence of this mass gap~\cite{Zevin:2020gma}. Continuing up the mass spectrum, a second mass gap is expected between roughly $65$--$120 \, M_\odot$ from pulsational pair-instability supernova theory~\cite{Woosley:2016hmi}. The location of the upper mass gap is subject to substantial uncertainty~\cite{Farmer:2019jed,Mapelli:2019ipt}, and the gap may be populated through dynamical formation channels~\cite{Kremer:2020wtp}. Even if the mass gap exists in the spectrum of newly formed stellar black holes, multi-generational mergers could populate the gap if the local stellar density is high enough~\cite{Rodriguez:2019huv,Gerosa:2021mno}. The detection of GW190521, with component masses of roughly $85 \, M_\odot$ and $66\, M_\odot$~\cite{LIGOScientific:2020iuh}, challenges the existence of the upper mass gap~\cite{LIGOScientific:2020ufj}, and the population analysis of the GWTC-3 performed by the LVK Collaboration found no evidence for a suppression of the merger rate above $60\, M_\odot$~\cite{KAGRA:2021duu}. Finally, beyond the upper limit of the upper mass gap, searches for merging binaries with component black holes in the Intermediate Mass Black Hole (IMBH) regime, have so far yielded negative results~\cite{LIGOScientific:2021tfm}. Well-fitting parametric ABH mass distributions are usually endowed with power law tails with upper mass cutoffs to represent this high-mass suppression, but the uncertainty in the formation mechanism of IMBHs makes a physical explanation challenging.

Whilst the distribution of chirp masses is the most constraining feature of the data for population studies, additional information comes from the distribution of less well-measured source parameters, in particular the spins, redshifts, and mass ratios of the binary systems. ABH models do not make strong predictions for the distributions of these quantities; instead, constraints from the GWTC provide key observational hints at the evolutionary pathways for BBH events. We summarise the key findings from these distributions in the ABH context below.

\begin{figure}[tbp!]
\centering
\includegraphics[width=\textwidth]{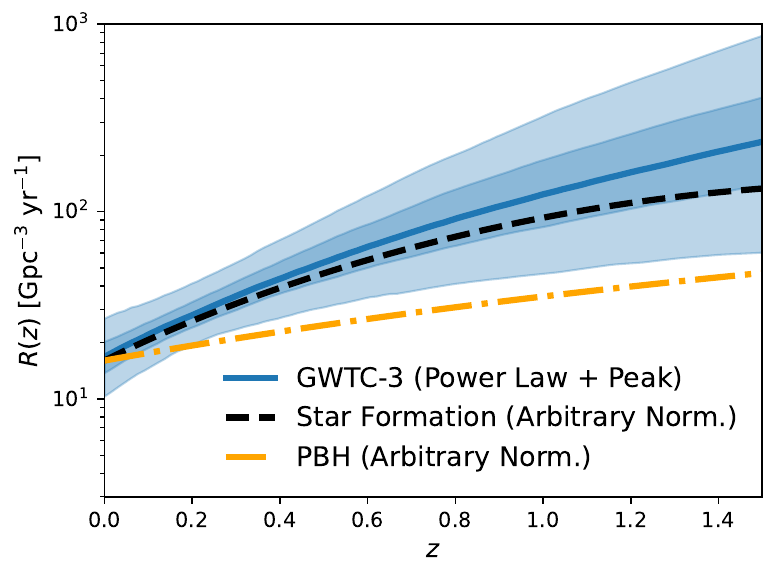}
\caption{Comoving merger rate density of BBH mergers in GWTC-3~\cite{KAGRA:2021duu}. We show the median (blue solid curve) and 50\% and 90\% credibility intervals (shaded regions). Overplotted is a fit to the global star formation rate density from Ref.~\cite{Madau:2014bja} (black dashed curve) and the PBH merger rate model from~\cite{Raidal:2018bbj} at fixed source mass and neglecting the suppression factor $S$ (orange dot-dashed curve), both with an arbitrary normalisation.}
\label{chap25:fig:rates}
\end{figure}

\begin{itemize}

\item Spin constraints are most succinctly summarised by the parameter $\chi_{\mathrm{eff}}$, the projection of a component mass-weighted total spin angular momentum vector onto the total angular momentum vector, which is typically measured with better precision than either spin component individually~\cite{Ajith:2009bn}. The LVK Collaboration has found evidence for a broadening of the distribution of $\chi_{\mathrm{eff}}$ above $30\, M_\odot$, and a negative correlation between spins and the mass ratio $q\equiv m_2/m_1$, where $m_1$ is the primary component (i.e.~the heavier of the two component masses), at 97.5\% credibility~\cite{KAGRA:2021duu}.

\item The redshift dependence of the BBH merger rate is well fit by a power law model having merger rate per comoving volume $\mathcal{R}(z) \propto (1+z)^\kappa$, with $\kappa = 2.9^{+1.7}_{-1.8}$ at 90\% credibility~\cite{KAGRA:2021duu}. Constraints on $\mathcal{R}(z)$ are summarized in Figure~\ref{chap25:fig:rates}. The redshift dependence is consistent with the Madau-Dickinson evolution of the global star formation rate~\cite{Madau:2014bja}.

\item Ref.~\cite{KAGRA:2021duu} finds that the distribution of mass ratio is well fitted by a power law in $q$ with exponent $\beta_q = 1.1^{+1.7}_{-1.3}$ at 90\% credibility. The constraints on $q$ are relatively broad compared with the prior for most events, making the mass ratio a relatively less powerful discriminator between models than proxies for the total mass. As is visible from Figure~\ref{chap25:fig:GWTC}, most of the systems in GWTC-3 have mass ratios compatible with unity, although it should be noted that extreme mass ratios ($q \ll 0.01$) are relatively less detectable for the LVK detector network at fixed total mass (see, e.g.~Figure 3 of Ref.~\cite{Hall:2020daa}).
\end{itemize}

In the context of constraining the possibility of a PBH population from GWTC-3, one should keep in mind that posterior constraints on CBC source parameters can be sensitive to the adopted priors. When constraining populations models, one effectively replaces these priors with a distribution for the parameters conditioned on the source population model. This model will typically have its own free parameters (``hyperparameters") that themselves require priors to be specified. As stressed in Ref.~\cite{Bhagwat:2020bzh}, the influence of the prior can strongly impact the source parameter posteriors for some GWTC events. Results from population-informed priors for empirical parametric ABH populations are presented alongside the default ``uninformative" priors in Ref.~\cite{KAGRA:2021duu}. The use of these alternative priors can impact, for example, conclusions drawn about the existence of the lower mass gap. The results discribed above and presented in Figures~\ref{chap25:fig:GWTC}, \ref{chap25:fig:mtot_kdes}, and \ref{chap25:fig:rates} assume the uninformative priors adopted in Ref.~\cite{KAGRA:2021duu}.

\subsection{PBH constraints assuming a single population}
\label{chap25:subsubsec:onepop}

We employ GW events from GWTC-3, excluding those with false alarm rates (FAR) exceeding 1 yr$^{-1}$ and secondary component masses below $3M_\odot$ to mitigate potential contamination from events involving neutron stars following Ref.~\cite{DeLuca:2021wjr}. This selection yields a total of $69$ GW events from GWTC-3 that satisfy these criteria, and the corresponding posterior samples for these BBH events are publicly accessible through Ref.~\cite{ligo_scientific_collaboration_and_virgo_2021_5655785}.
In the analyses, we use combined posterior samples obtained from the IMRPhenom~\cite{Thompson:2020nei,Pratten:2020ceb} and SEOBNR~\cite{Ossokine:2020kjp,Matas:2020wab} waveform families.

\begin{table}[tbp!]
\centering
\begin{adjustbox}{width=0.6\textwidth}
\begin{tabular}{lll}
\hline
\textbf{Parameter\quad} & \textbf{Description} & \textbf{Prior} \\
\hline
$f_\mathrm{PBH}$ & Abundance of PBH in CDM. & log$\mathcal{U}(-4, 0)$\\
\hline
\multicolumn{3}{c}{Lognormal PBH mass function} \\[1pt]
$M_\mathrm{c}$ & Central mass in $M_\odot$. & $\mathcal{U}(5, 50)$\\
$\sigma$ & Mass width. & $\mathcal{U}(0.1, 2)$\\
\hline
\multicolumn{3}{c}{Power-law PBH mass function} \\[1pt]
$M_\mathrm{min}$ & Lower mass cutoff in $M_\odot$. & $\mathcal{U}(3, 10)$\\
$\alpha$ & Power-law index. & $\mathcal{U}(1.05, 4)$\\
\hline
\multicolumn{3}{c}{Broken Power-law (BPL) PBH mass function} \\[1pt]
$m_*$ & Peak mass in $M_\odot$. & $\mathcal{U}(5, 50)$\\
$\alpha_1$ & First power-law index. & $\mathcal{U}(0, 3)$\\
$\alpha_2$ & Second power-law index. & $\mathcal{U}(1, 10)$\\
\hline
\multicolumn{3}{c}{Critical collapse (CC) PBH mass function} \\[1pt]
$M_\mathrm{f}$ & Horizon mass scale in $M_\odot$. & $\mathcal{U}(1, 50)$\\
$\alpha$ & Universal exponent. & $\mathcal{U}(0, 5)$\\
\hline
\end{tabular}	
\end{adjustbox}
\caption{Model parameters and their corresponding prior distributions employed in our Bayesian parameter estimation for the four PBH mass function models. Here, $\mathcal{U}$ and log$\mathcal{U}$ refer to uniform and log-uniform distributions, respectively.}
\label{chap25:tab:prior_pbh}
\end{table}

\begin{figure}[tbp!]
\centering
\includegraphics[width=0.48\textwidth]{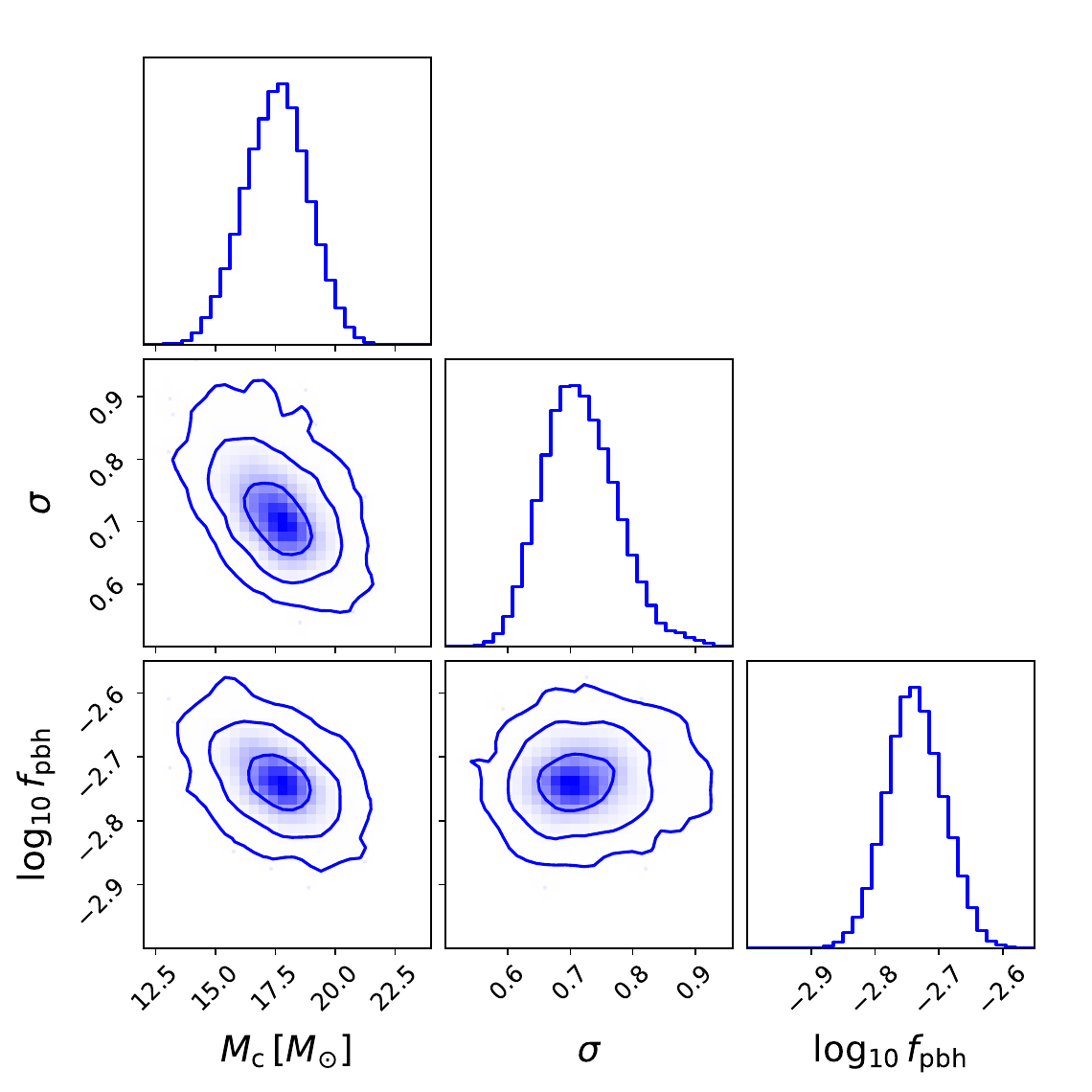}
\includegraphics[width=0.48\textwidth]{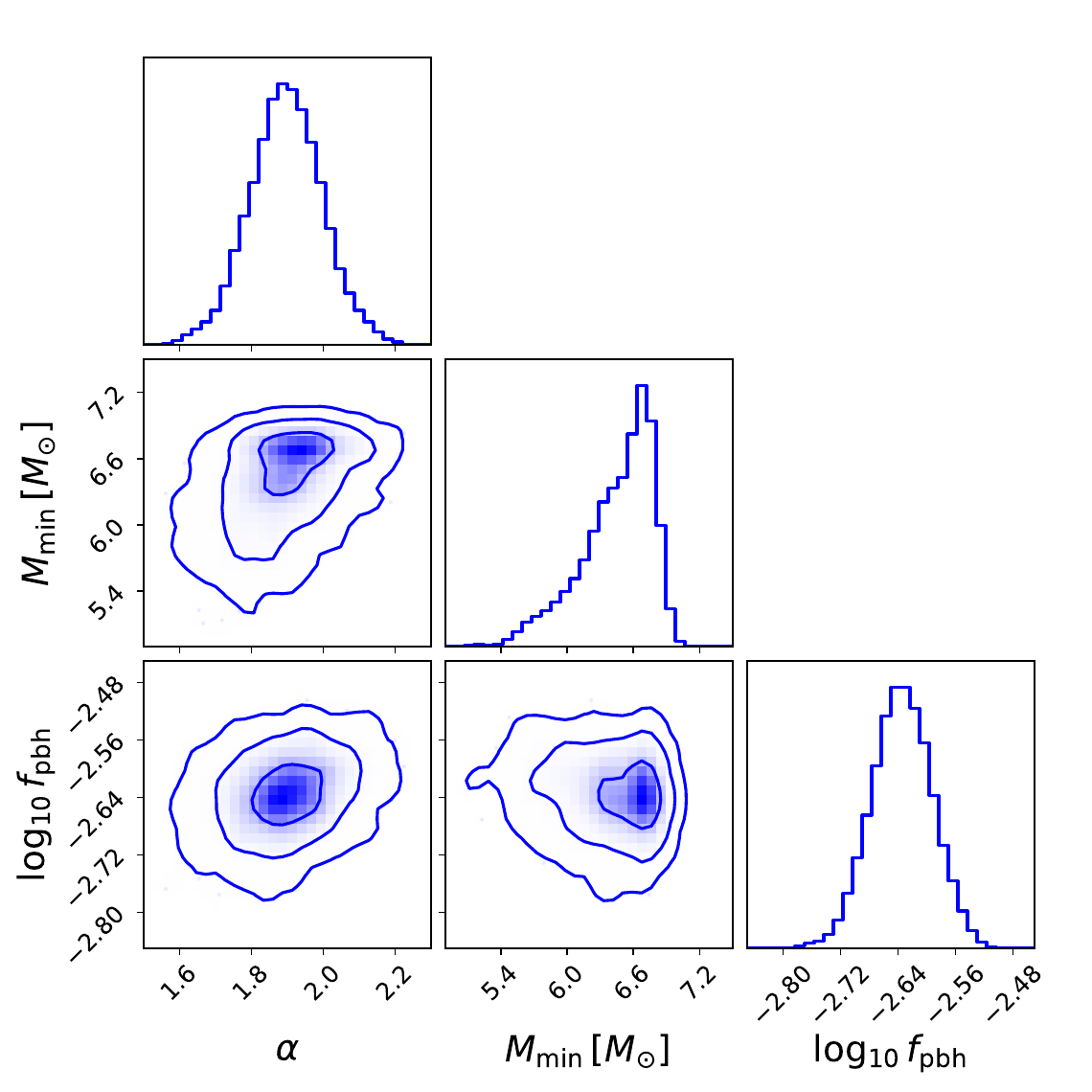}\\
\includegraphics[width=0.48\textwidth]{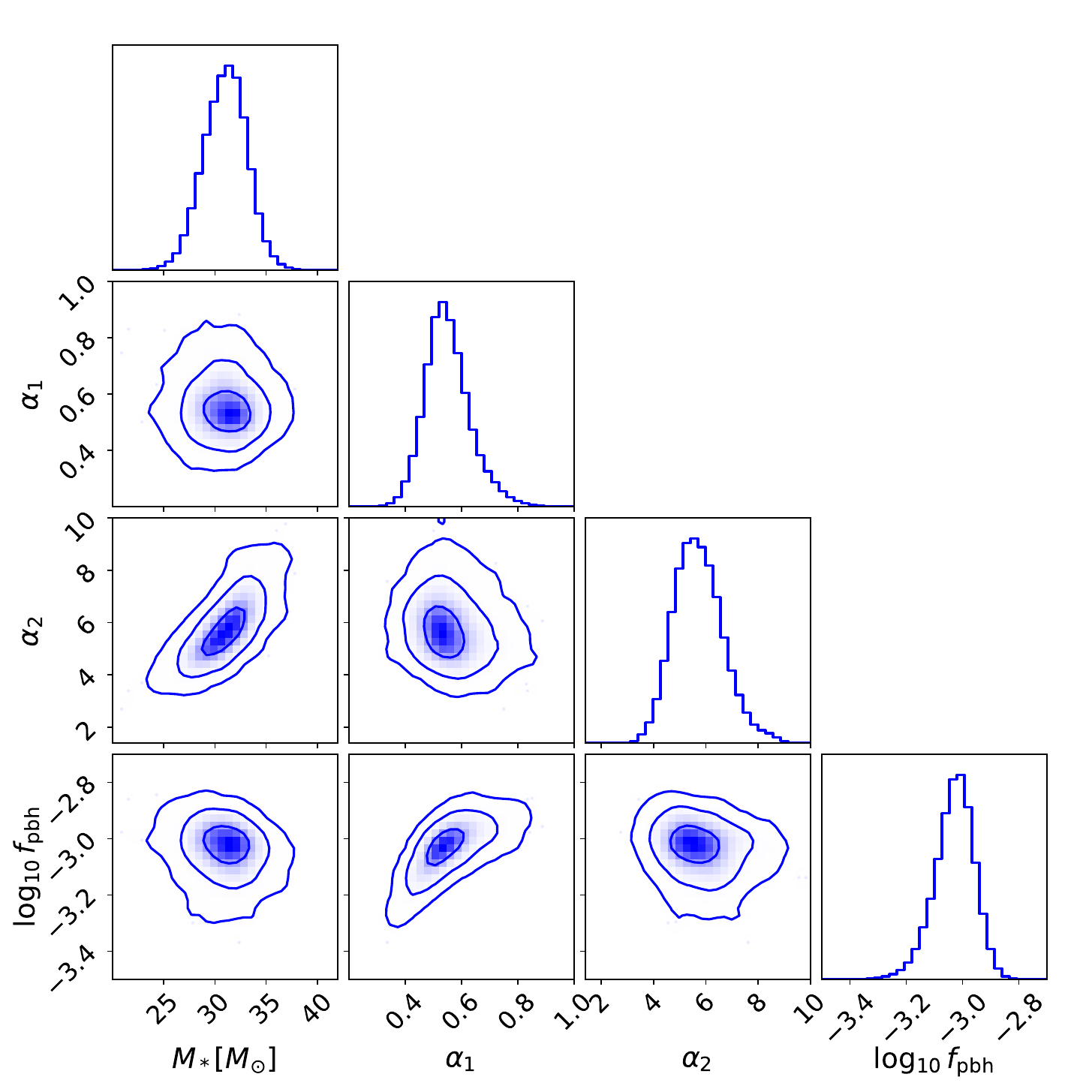}
\includegraphics[width=0.48\textwidth]{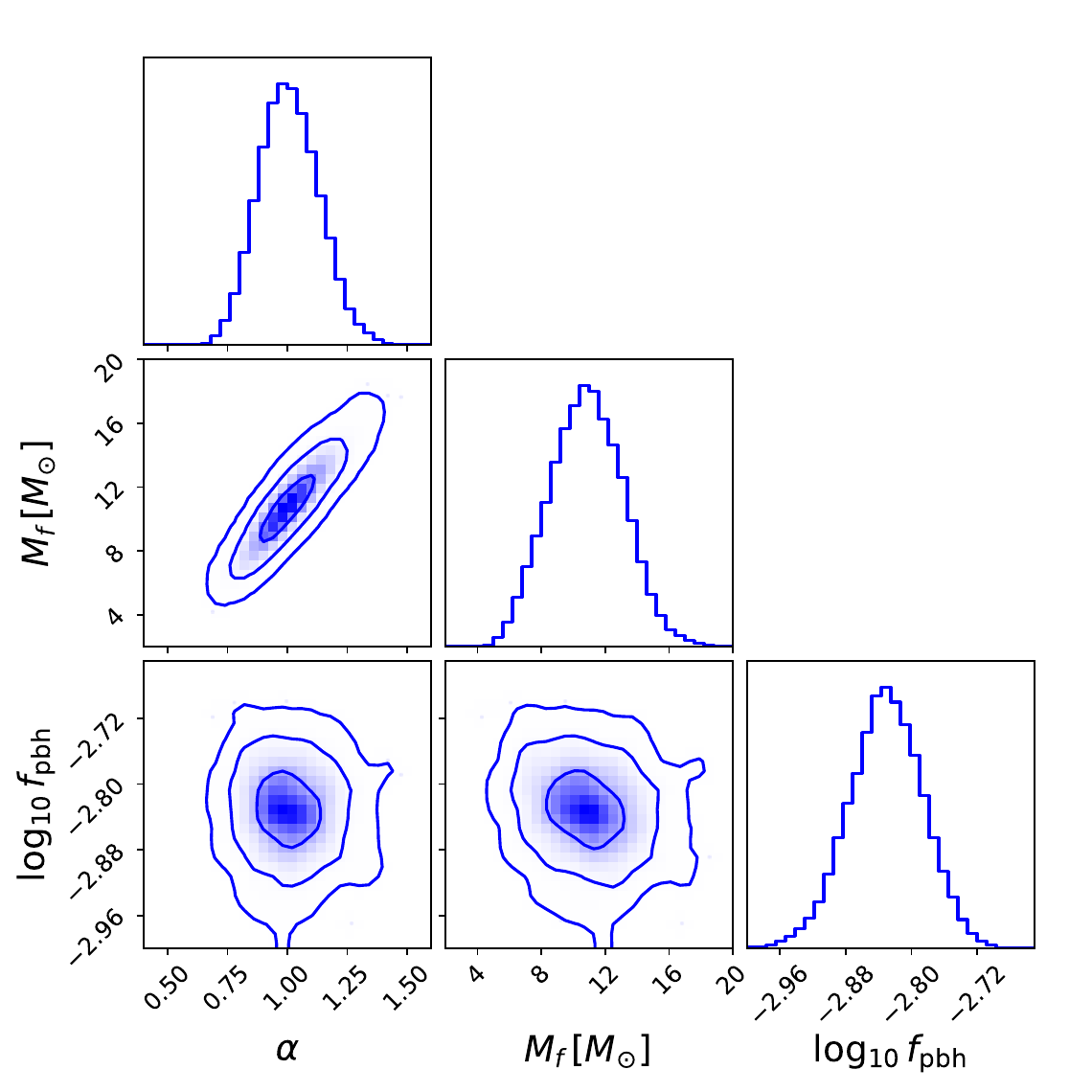}
\caption{
Marginalized posterior distributions for hyperparameters inferred from GWTC-3, with each panel corresponding to a specific mass function: lognormal (top left), power-law (top right), broken power-law (bottom left), and critical collapse (bottom right). Contours denote the $1\sigma$, $2\sigma$, and $3\sigma$ credible regions, respectively.}
\label{chap25:fig:posterior_pbh}
\end{figure}

\begin{figure}[tbp!]
\centering
\includegraphics[width=\textwidth]{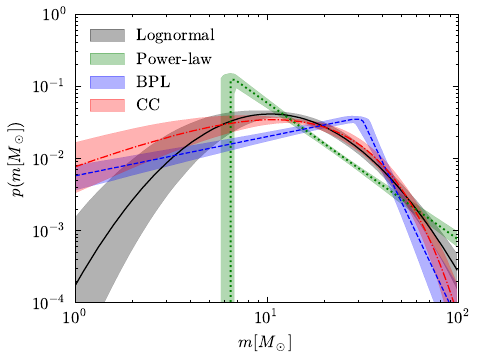}
\caption{Posterior predictive distributions for the four PBH mass functions using GWTC-3 catalog.}
\label{chap25:fig:ppd_PBH}
\end{figure}

Employing the Bayesian inference framework outlined in Section~\ref{chap25:subsec:bayes}, we perform parameter estimations for four distinct PBH mass functions, as described in Section~\ref{chap25:subsec:mergerrate}. A comprehensive summary of the parameters and their corresponding prior distributions is provided in Table~\ref{chap25:tab:prior_pbh}. Utilizing data from $69$ BBH events from GWTC-3 and employing hierarchical Bayesian inference, we obtain posterior distributions for the hyperparameters, as illustrated in Fig.~\ref{chap25:fig:posterior_pbh}. Additionally, the posterior predictive distributions (PPD) for the four PBH mass functions are depicted in Fig.~\ref{chap25:fig:ppd_PBH}.

The PPD updates the prior for parameters $\theta$ based on data $\mathbf{d}$. Recall the hyper-posterior $p(\Lambda | \mathbf{d})$ reflects our post-measurement understanding of hyperparameters shaping the mass distribution $P(\theta)$. The PPD addresses: given this hyper-posterior, what is the distribution of $P(\theta)$? It signifies the probability that the next event possesses true parameter values $\theta$ based on our knowledge of population hyperparameters $\Lambda$:
\begin{equation}
p_\Lambda(\theta | \mathbf{d})=\int \mathrm{d} \Lambda\, p(\Lambda | \mathbf{d})\, P(\theta | \Lambda).    
\end{equation}
The subscript $\Lambda$ distinguishes the PPD from the posterior $p(\theta | \mathbf{d})$. In the hyper-posterior sample version, the PPD is expressed as
\begin{equation}
p_{\Lambda}(\theta | \mathbf{d})=\frac{1}{n_s} \sum_k^{n_s} P(\theta | \Lambda_k),
\end{equation}
where $k$ iterates over $n_s$ hyper-posterior samples. While the PPD best estimates the appearance of $P(\theta)$, it doesn't convey potential variability in $P(\theta)$ due to uncertainty in $\Lambda$. Effectively conveying this variability may involve overplotting multiple realizations of $P(\theta|\Lambda_k)$, where $\Lambda_k$ is a randomly chosen hyper-posterior sample.

For the lognormal PBH mass function model, our analysis yields hyperparameter values with median estimates and $90\%$ equal-tailed credible intervals: $M_c = 17.3^{+2.2}_{-2.0} M_\odot$, $\sigma = 0.71^{+0.10}_{-0.08}$, and $f_\mathrm{PBH} = 1.8^{+0.3}_{-0.3} \times 10^{-3}$.  
It's noteworthy that the derived $M_c$ value is larger than that obtained from GWTC-1 in Ref.~\cite{Wu:2020drm} due to the inclusion of heavier BHs in GWTC-3. Moreover, utilizing Eq.~\eqref{chap25:eq:Rt}, we deduce the local merger rate as $R(t_0) = 41^{+16}_{-12} \mathrm{Gpc}^{-3}\,\mathrm{yr}^{-1}$.

For the power-law PBH mass function model, our analysis yields $M_\mathrm{min} = 6.5^{+0.3}_{-0.8} M_\odot$, $\alpha = 1.9^{+0.2}_{-0.2}$, and $f_\mathrm{PBH} = 2.3^{+0.3}_{-0.3} \times 10^{-3}$. 
Notably, the inferred $\alpha$ value is smaller than that from GWTC-1 in Ref.~\cite{Wu:2020drm} due to the inclusion of more massive BHs in GWTC-3. Using Eq.~\eqref{chap25:eq:Rt}, we estimate the local merger rate as $R(t_0) = 48^{+15}_{-12} \mathrm{Gpc}^{-3},\mathrm{yr}^{-1}$.

For the broken power-law PBH mass function model, our analysis yields $m_* = 31.1^{+1.8}_{-2.1} M_\odot$, $\alpha_1 = 0.54^{+0.08}_{-0.06}$, $\alpha_2 = 5.6^{+0.9}_{-0.8}$, and $f_\mathrm{PBH} = 0.9^{+0.1}_{-0.1} \times 10^{-3}$.
Applying Eq.~\eqref{chap25:eq:Rt}, we estimate the local merger rate as $R(t_0) = 46^{+15}_{-11} \mathrm{Gpc}^{-3}\,\mathrm{yr}^{-1}$.

For the critical collapse PBH mass function model, our analysis yields $M_\mathrm{f} = 10.8^{+3.7}_{-3.6} M_\odot$, $\alpha = 1.0^{+0.2}_{-0.2}$, and $f_\mathrm{PBH} = 1.5^{+0.2}_{-0.2} \times 10^{-3}$. 
Using Eq.~\eqref{chap25:eq:Rt}, we estimate the local merger rate as $R(t_0) = 49^{+26}_{-16} \mathrm{Gpc}^{-3}\,\mathrm{yr}^{-1}$.

Across all four mass functions, our inferences suggest that the abundance of PBHs in CDM, $f_\mathrm{PBH}$, is on the order of $\mathcal{O}(10^{-3})$. These results align with previous estimations~\cite{Sasaki:2016jop,Ali-Haimoud:2017rtz,Chen:2018czv,Chen:2018rzo,Chen:2019irf,Wu:2020drm,Chen:2021nxo,Chen:2022fda,Zheng:2022wqo}, supporting the conclusion that stellar-mass PBHs do not dominate the composition of CDM. While the late-time clustering of PBHs may impact the observed merger rate, our inferred values for $f_{\mathrm{PBH}}$ consistently remain below $3 \times 10^{-3}$ for all four PBH mass functions. Therefore, this clustering effect can be safely neglected according to Ref.~\cite{Hutsi:2020sol}.

\begin{table}[tbp!]
\centering
\begin{adjustbox}{width=0.5\textwidth}
\begin{tabular}{c|c|c|c|c}
\hline
& lognormal & power-law & BPL & CC \\
\hline
$\mathrm{BF}_{\mathrm{PL}}$ & $166$ & $1$ & $49$ & $139$ \\
\hline
\end{tabular}
\end{adjustbox}
\caption{Bayes factors, $\mathrm{BF}_{\mathrm{PL}}$, comparing models with various PBH mass functions to the model with a power-law PBH mass function.}
\label{chap25:tab:BF_PBH}
\end{table}

We also calculate the Bayes factors ($\mathrm{BF}_{\mathrm{PL}}$) between models with different PBH mass functions, using the model with the power-law mass function as the fiducial model. From Table~\ref{chap25:tab:BF_PBH}, we see that $\mathrm{BF}_{\mathrm{PL}}^{\mathrm{LG}}$ attains the highest value, suggesting that the lognormal mass function provides the most optimal fit to GWTC-3 among the four mass functions. 

While the results presented in this subsection suggest a lognormal or critical-collapse mass function is preferred over power law or broken power law forms, it is important to note that none of the considered mass functions provides a particularly good fit to the observed BBH mass distribution. As discussed in Section~\ref{chap25:subsubsec:ABH} and emphasized in Refs.~\cite{Hall:2020daa,Franciolini:2022tfm}, the observed distribution has features, such as multiple peaks, that are not captured by any of our assumed PBH mass functions. This fact presents a stark challenge to the PBH scenario in the BBH context, and motivates the consideration of mixed ABH+PBH populations, to which we now turn.

\subsection{PBH constraints assuming mixed populations}
\label{chap25:subsubsec:mixedpop}

Previous investigations suggest the possibility that the BBH mergers observed by the LVK Collaboration may originate from both ABHs and PBHs~\cite{Hall:2020daa,Hutsi:2020sol,Wong:2020yig,DeLuca:2021wjr,Franciolini:2021tla,Franciolini:2022tfm}. In this section, we explore a hybrid model, referred to as the mixed ABH+PBH model, wherein BBHs can arise from both ABH and PBH channels.

\begin{table}[htbp!]
\centering
\begin{adjustbox}{width=\textwidth}
\begin{tabular}{lll}
\hline
\textbf{Parameter} & \textbf{Description} & \textbf{Prior} \\
\hline
$\gamma$ & Slope of the power-law regime for the rate evolution before the point $z_p$. & $\mathcal{U}(0, 12)$\\
$\kappa$ & Slope of the power-law regime for the rate evolution after the point $z_p$. & $\mathcal{U}(0, 6)$\\
$z_p$ & Redshift turning point between the power-law regimes with $\gamma$ and $\kappa$. & $\mathcal{U}(0, 4)$\\
$\alpha$ & Spectral index for the power-law of the primary mass distribution. & $\mathcal{U}(1.5, 12)$\\
$\beta$ & Spectral index for the power-law of the mass ratio distribution. & $\mathcal{U}(-4, 12)$\\
$m_{\min}\,[M_\odot]$ & Minimum mass of the power-law component of the primary mass distribution. & $\mathcal{U}(2, 10)$\\
$m_{\max}\,[M_\odot]$ & Maximum mass of the power-law component of the primary mass distribution. & $\mathcal{U}(50, 200)$\\
$\lambda_g$ & Fraction of the model in the Gaussian component. & $\mathcal{U}(0, 1)$\\
$\mu_g\,[M_\odot]$ & Mean of the Gaussian component in the primary mass distribution. & $\mathcal{U}(20, 50)$\\
$\sigma_g\,[M_\odot]$ & Width of the Gaussian component in the primary mass distribution. & $\mathcal{U}(0.4, 10)$\\
$\delta_m\,[M_\odot]$ & Range of mass tapering at the lower end of the mass distribution. & $\mathcal{U}(0, 10)$\\
\hline
\end{tabular}
\end{adjustbox}
\caption{Model parameters and their corresponding prior distributions employed in our Bayesian parameter estimation for the ABH model.}
\label{chap25:tab:prior_abh}
\end{table}

\begin{figure}[tbp!]
\centering	
\includegraphics[width=\textwidth]{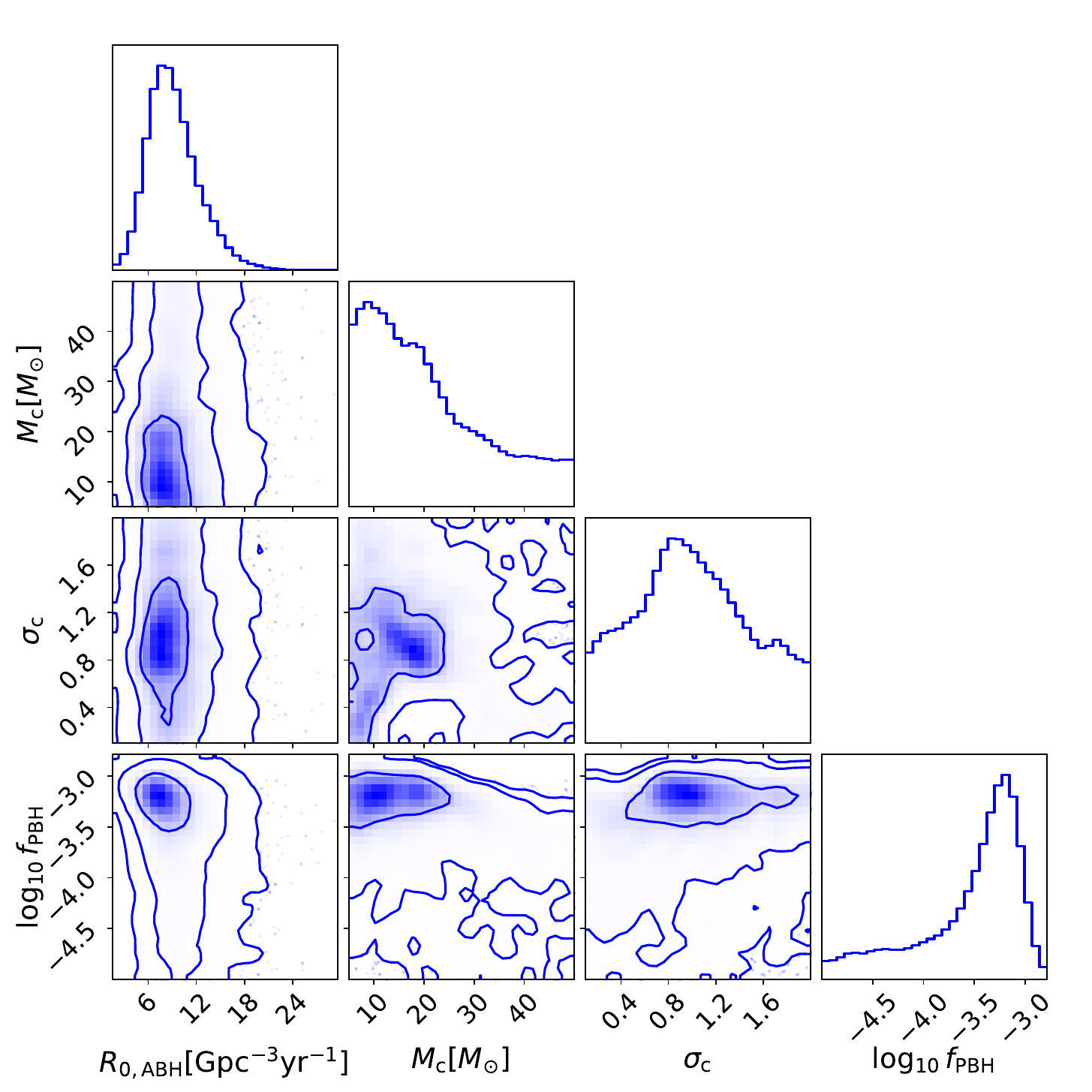}
\caption{One and two-dimensional marginalized posteriors for some of the hyperparameters in the mixed ABH+PBH model.}
\label{chap25:fig:post_abh_pbh}
\end{figure}

We adopt the lognormal mass function for the PBH component, as it is a reasonable fit to the mass function resulting from a peak in the primordial curvature perturbation. For the ABH component, we employ a phenomenological model following Ref.~\cite{LIGOScientific:2021aug}. The combined merger rate, $\mathcal{R}_{\mathrm{total}}(m_1, m_2, z)$, is the sum of the ABH merger rate ($\mathcal{R}_{\mathrm{ABH}}$) and the PBH merger rate ($\mathcal{R}_{\mathrm{PBH}}$), given by
\begin{equation}
	\mathcal{R}_{\mathrm{total}}(z, m_1, m_2) = \mathcal{R}_{\mathrm{ABH}}(z, m_1, m_2) + \mathcal{R}_{\mathrm{PBH}}(z, m_1, m_2),
\end{equation}
where $\mathcal{R}_{\mathrm{PBH}}(z, m_1, m_2)$ is given by Eq.~\eqref{chap25:eq:cR}. Specifically, the ABH merger rate is expressed as
\begin{equation}
	\mathcal{R}_{\mathrm{ABH}}(z, m_1, m_2) = R_{0, \mathrm{ABH}} \, \pi(z) \, \pi(m_1)\, \pi(m_2| m_1),
\end{equation}
where $R_{0, \mathrm{ABH}}$ represents the local merger rate of ABH binaries. 

The redshift distribution $\pi(z)$ is parameterized  as
\begin{equation}
\label{chap25:eq:pz}
\pi(z| \gamma, \kappa, z_{\mathrm{p}}) = \left[1+\left(1+z_{\mathrm{p}}\right)^{-\gamma-\kappa}\right]  \frac{(1+z)^\gamma}{1+\left[(1+z) / (1+z_\mathrm{p})\right]^{\gamma + \kappa}},
\end{equation}
where $\gamma$ and $k$ denote the slopes of two power-law regimes before and after a turning point $z_{\mathrm{p}}$. This choice of parameterization is motivated by the potential correlation between the binary formation rate and the star formation rate~\cite{Madau:2014bja,Callister:2020arv}.

The primary mass distribution $\pi(m_1)\equiv\pi\left(m_1| m_{\min }, m_{\max }, \alpha, \lambda_{\mathrm{g}}, \mu_{\mathrm{g}}, \sigma_{\mathrm{g}}\right)$ is chosen to follow the ``Power Law + Peak" model~\cite{KAGRA:2021duu}, which is a combination of a power-law and Gaussian component, given by
\begin{equation}	
	\pi(m_1)=\left[(1-\lambda_{\mathrm{g}}) \mathcal{P} (m_1| m_{\min }, m_{\max },-\alpha)+\lambda_{\mathrm{g}} \mathcal{G}(m_1 | \mu_{\mathrm{g}}, \sigma_{\mathrm{g}})\right] S(m_1, m_{\min }, \delta_m),
\end{equation}
where $\mathcal{P}(x | x_{\min}, x_{\max}, \alpha)$ represents a truncated power-law with slope $\alpha$ between the bounds $x_{\min }$ and $x_{\max}$, given by
\begin{equation}
	\mathcal{P}(x | x_{\min }, x_{\max }, \alpha) \propto 
 \begin{cases}
 x^\alpha & \left(x_{\min } \leqslant x \leqslant x_{\max }\right) \\ 0 & \text{otherwise.}
 \end{cases}
\end{equation}
The Gaussian distribution, $\mathcal{G}(x | \mu, \sigma)$, is characterized by mean $\mu$ and standard deviation $\sigma$:
\begin{equation}
	\mathcal{G}(x | \mu, \sigma)=\frac{1}{\sigma \sqrt{2 \pi}} \exp \left[-\frac{(x-\mu)^2}{2 \sigma^2}\right].
\end{equation}

The parameter $\lambda_{\mathrm{g}}$ serves as the ratio parameter between the power-law component $\mathcal{P}$ and the Gaussian component $\mathcal{G}$. Additionally, the sigmoid-like window function, $S(m_1, m_{\text {min }}, \delta_m)$, introduces a smoothing rise within the interval $\left(m_{\min }, m_{\min }+\delta_m\right)$:
\begin{equation}
	S(m, m_{\min }, \delta_m)= \begin{cases}0 & \left(m<m_{\min }\right) \\ {\left[f\left(m-m_{\min }, \delta_m\right)+1\right]^{-1}} & \left(m_{\min } \leq m<m_{\min }+\delta_m\right) \\ 1 & \left(m \geq m_{\min }+\delta_m\right),\end{cases}
\end{equation}
where $f(m^{\prime}, \delta_m)$ is defined by
\begin{equation}
	f(m^{\prime}, \delta_m)=\exp \left(\frac{\delta_m}{m^{\prime}}+\frac{\delta_m}{m^{\prime}-\delta_m}\right).
\end{equation}

The secondary mass distribution $\pi(m_2)$ is modeled with a truncated power-law with slope $\beta$ between a minimum mass $m_{\min}$ and the primary mass $m_1$ as
\begin{equation}
	\pi(m_2 | m_1, m_{\min }, \alpha)=\mathcal{P}(m_2 | m_{\min }, m_1, \beta)\, S(m_2, m_{\min }, \delta_m).
\end{equation}

The priors for the parameters associated with the lognormal PBH mass function align with those presented in Table~\ref{chap25:tab:prior_pbh}. Concurrently, the priors for parameters in the ABH model are outlined in Table~\ref{chap25:tab:prior_abh}. Using the data from $69$ BBHs obtained from GWTC-3 and employing hierarchical Bayesian inference, we derive the posteriors for the hyperparameters within the mixed ABH+PBH model, as illustrated in Fig.~\ref{chap25:fig:post_abh_pbh}.

Our analysis yields hyperparameter values with median estimates and $90\%$ equal-tailed credible intervals: $M_c = 18.1^{+26.7}_{-11.6} M_\odot$, $\sigma = 0.98^{+0.87}_{-0.75}$, and $f_\mathrm{PBH} = 4.4^{+5.3}_{-4.1} \times 10^{-4}$. The inclusion of the ABH channel introduces notably larger uncertainties in the inferred values for $M_c$ and $\sigma$, with a substantial reduction of the upper limit on $f_\mathrm{PBH}$, as expected. Additionally, we determine the local merger rate for ABHs as $R_{0, \mathrm{ABH}} = 8.8^{+6.2}_{-3.7} \mathrm{Gpc}^{-3} \mathrm{yr}^{-1}$.

Within the mixed ABH+PBH model, our analysis indicates that the fraction of detectable events originating from PBH binaries in GWTC-3 is $f_{\rm P}\equiv N^{\rm det}_{\rm PBH}/(N^{\rm det}_{\rm PBH}+N^{\rm det}_{\rm ABH}) = 24.5^{+30.6}_{-17.3}\%$. This finding is consistent with the results obtained in previous studies~\cite{DeLuca:2021wjr,Zheng:2022wqo}. Despite the substantial uncertainty associated with $f_{\rm P}$, it implies that at least a subset of BBHs in GWTC-3 can be attributed to the PBH channel.

So far, we have parameterised the PBH population in terms of its mass function. An alternative approach, demonstrated in Refs.~\cite{Zheng:2022wqo,Franciolini:2022tfm}, is to drill down further and parameterise the curvature power spectrum that gives rise to PBHs. This approach is a potentially powerful way to constrain inflationary models directly from GW source catalogs, but will require further progress in developing accurate models of the ABH binary population to realize its full potential.

To conclude this subsection, we have seen that GW data are not consistent with models in which all of the sources are PBH binaries. Such models are unable to reproduce the detailed features of the black hole mass distribution that GWTC-3 has revealed. Mixed populations on the other hand do allow for a non-zero PBH contribution to the observed population, but this result should be treated with extreme caution due to the large uncertainties associated with ABH population models. While a PBH component could explain the presence of objects in the ABH mass gaps, such as GW190814, it remains to be seen whether the existence of these gaps is a strong enough prediction of stellar physics that PBHs can be considered a leading candidate for the observed GW sources.

\section{Prospects for constraints with the Einstein Telescope}
\label{chap25:sec:future}

The second-generation GW detectors, Advanced LIGO and Advanced Virgo, have significantly advanced our understanding of the Universe. Despite their ongoing improvements in sensitivity, these detectors are anticipated to encounter limitations in the future, primarily due to constraints imposed by their hosting infrastructures. Consequently, the exploration of new GW detectors is underway to overcome these limitations. One such initiative is the Einstein Telescope (ET)~\cite{Hild:2008ng,Punturo:2010zz,Hild:2010id}, envisioned as a third-generation GW observatory in Europe. ET represents a novel research infrastructure designed to accommodate a detector with the capability to observe the entire Universe through GWs. It is conceptualized as a multi-interferometer observatory with the ambitious goal of enhancing sensitivity by a factor of ten compared to previous-generation detectors. See chapter~16 for an overview of various GW detectors.

This section delves into the exploration and projection of constraints on PBHs and the potential to distinguish them from ABHs using third-generation ground-based GW detectors such as ET. These advanced detectors are expected to surpass the capabilities of the current LVK detector network, potentially detecting on the order of $\mathcal{O}(10^5)$ BBH events annually~\cite{Regimbau:2016ike,Vitale:2018yhm}.

\subsection{Constraints on the PBH abundance from the ET}
\label{chap25:subsec:constraintET}

The detection of PBHs serves to constrain the PBH abundance parameter $f_\mathrm{PBH}$. Conversely, the absence of PBH detection will place upper limits on the merger rate of PBH binaries, consequently constraining the PBH abundance.
The detectable event rate of PBH binaries hinges on both the merger rate and the sensitivity of GW detectors. In this subsection, under the assumption of PBHs having a monochromatic mass distribution, we proceed to estimate the detectable limits on the abundance of PBHs through a targeted search conducted by ET.

The expected number of detections, $N_\mathrm{exp}$, can be estimated by~\cite{Chen:2018czv,Kavanagh:2018ggo}
\begin{equation}
\label{chap25:eq:Nobs} 
	N_\mathrm{exp} = \int R(z) \frac{\mathrm{d} VT}{\mathrm{d} z} \mathrm{d} z,
\end{equation}
where $\mathrm{d} VT/\mathrm{d} z$ characterizes the spacetime sensitivity of a GW detector as a function of redshift and incorporates the selection effects inherent to that particular detector. See also Eq.~\eqref{chap25:eq:L1}.
Generally, $\mathrm{d} VT/\mathrm{d} z$ depends on the properties $\theta$ (e.g. masses and spin) of a binary and is defined as~\cite{Abbott:2016nhf,Abbott:2016drs}
\begin{equation}
\label{chap25:eq:dVTdz}
\frac{\mathrm{d} VT}{\mathrm{d} z} = \frac{\mathrm{d} V_c}{\mathrm{d} z} \frac{T_\mathrm{obs}}{1+z} P_{\mathrm{det}}(\theta),   
\end{equation}
where $T_\mathrm{obs}$ is the observing time, and the denominator $1+z$ accounts for the converting of cosmic time from source frame to detector frame due to cosmic redshift.

\begin{figure}[tbp!]
\centering
\includegraphics[width=1\textwidth]{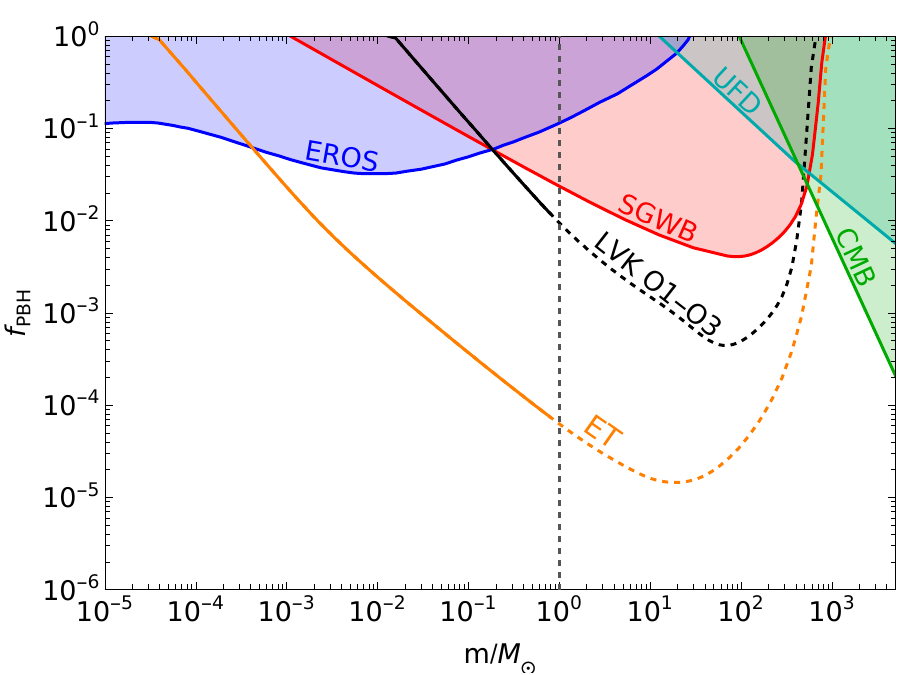}
\caption{Constraints on the PBH abundance, $f_\mathrm{PBH}$, characterized by a monochromatic mass distribution. These constraints are from both the non-detection of SGWBs during LVK O1-O3 runs and the null result of targeted searches for PBH systems. The vertical gray line at $1 M_\odot$ signifies that the constraints from the targeted search are applicable solely to sub-solar mass PBHs, as it remains uncertain whether any of the BBHs detected by LVC are composed of PBH binaries. The black and orange curves are the results of the targeted search from the LVK O1-O3 runs and the ET, respectively, assuming an observation time of one year for ET. The red curve represents the updated upper bound of $f_\mathrm{PBH}$ constrained by the non-detection of SGWBs in LIGO O1-O3 searches. Additionally, results from other experiments are included: EROS/MACHO microlensing (EROS)~\cite{Tisserand:2006zx}, dynamical heating of ultra-faint dwarf galaxies (UFD)~\cite{Brandt:2016aco}, and accretion constraints derived from CMB observations~\cite{Ali-Haimoud:2016mbv,Blum:2016cjs,Horowitz:2016lib,Chen:2016pud,Poulin:2017bwe}.}
\label{chap25:fig:fpbh_ET}
\end{figure}

Considering a scenario in which ET does not detect any PBH binaries, the observatory will establish an upper limit on the merger rate of PBH binaries. The $90\%$ credible level upper limit on the binary merger rate is determined using the loudest event statistic formalism~\cite{Biswas:2007ni}:
\begin{equation}
\label{chap25:eq:R90} 
R_{90} = \frac{2.303}{VT},
\end{equation}
where 
\begin{equation}
\label{chap25:eq:VT}
VT = \int \frac{\mathrm{d} VT}{\mathrm{d} z} \mathrm{d} z.
\end{equation}
To compute $VT$, we employ the semi-analytical approximation from~\cite{Abbott:2016nhf,Abbott:2016drs}, neglecting the effect of spins for BHs and utilizing the \texttt{IMRPhenomPv2} waveform to simulate the BBH templates. Additionally, we set a single-detector SNR threshold $\rho_{\mathrm{th}} = 8$ as a detection criterion, corresponding approximately to a network threshold of $12$.

The detectable limits for $f_\mathrm{PBH}$ through the targeted search using data from LVK's first three runs (O1-O3) and ET are illustrated in Fig.~\ref{chap25:fig:fpbh_ET}. Our assumptions include the absence of viable PBH binary candidates during ET observations, and ET operating at full duty for one year. Given the ongoing debate regarding whether the observed super-solar mass BBHs are of PBH origin or not, we represent the detectable limits for super-solar mass PBHs with dashed lines. Notably, if no PBH binaries are detected, ET is anticipated to constrain $f_\mathrm{PBH}$ at the order of $\mathcal{O}(10^{-4})$ for $1\, M_\odot$, which is two orders of magnitude more stringent than that from LVK O1-O3.

\subsection{Distinguishing PBHs from ABHs}
\label{chap25:subsec:distinguishET}

Currently, the second-generation GW detectors, such as LIGO, Virgo, and KAGRA, are limited to observing BBHs at low redshifts ($z < 1$), but future third-generation GW detectors like ET will extend the reach to much higher redshifts ($z \ge 10$). In this subsection, we will explore how ET, can effectively distinguish between PBHs and ABHs based on their distinct event rate distributions at high redshifts.

We have seen in Eq.~\eqref{chap25:eq:cR} that the merger rate of PBH binaries increases with redshift $z$, specifically $\mathrm{R}(z) \propto t(z)^{-34/37}$~\cite{Chen:2018czv}.
This relationship is independent of both the abundance and mass function of PBHs. In contrast, the merger rate of ABH binaries is thought to mimic the global star formation rate history; an initially rise with $z$, peaking at a low redshift, and then rapidly declining as described by Eq.~\eqref{chap25:eq:pz}.

\begin{figure}[tbp!]
\centering
\includegraphics[width=0.48\textwidth]{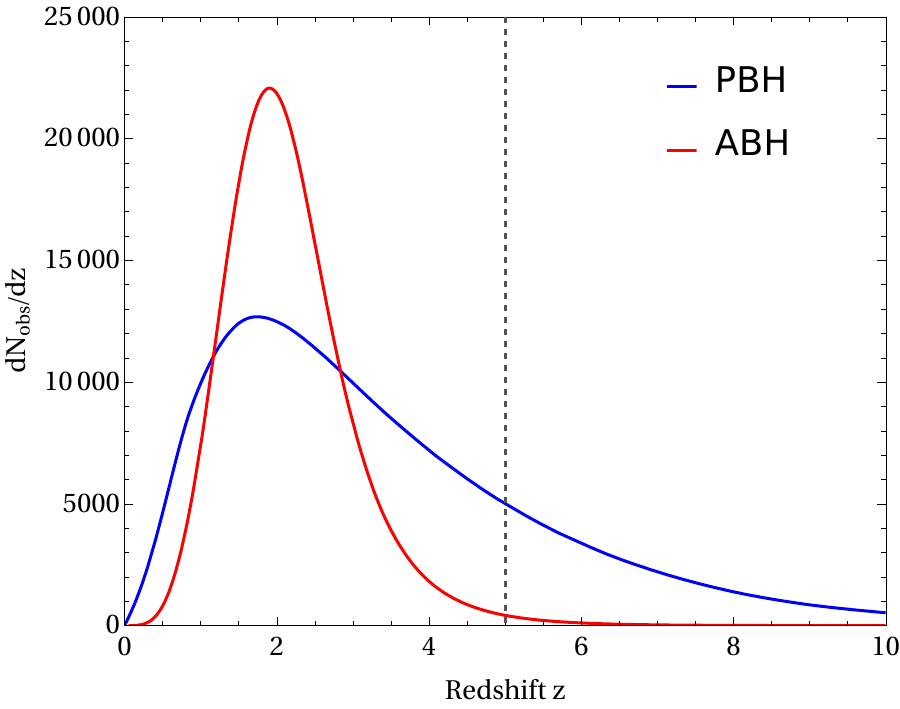}
\includegraphics[width=0.48\textwidth]{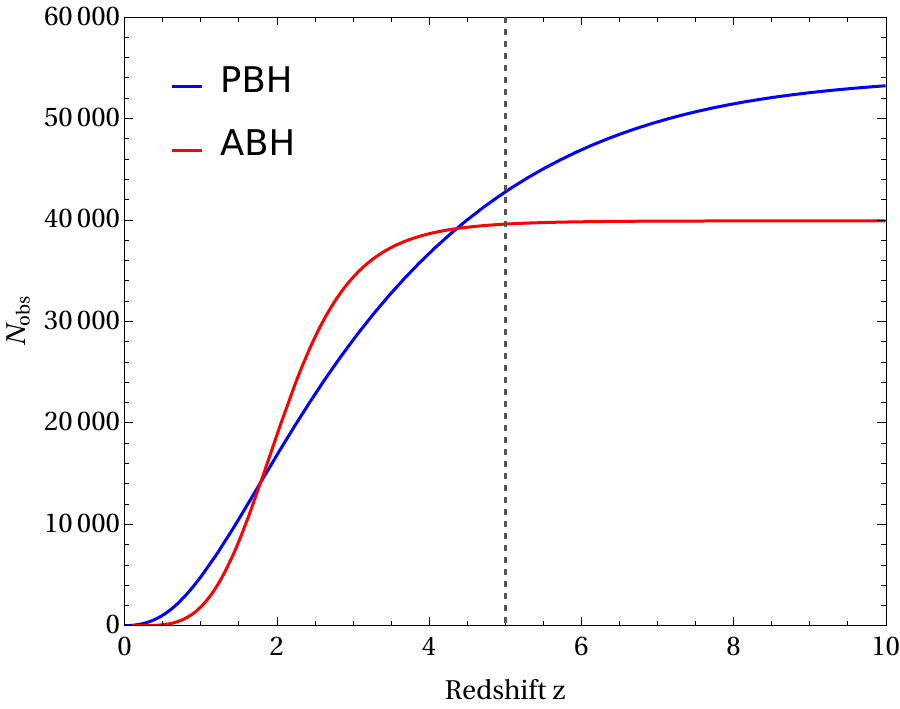}
\caption{\textbf{Left panel}: Redshift distribution of the expected number of observable BBHs, $\mathrm{d} N_\mathrm{exp}/\mathrm{d} z$, for ET. We have fixed the model parameters to their best-fit values from section~\ref{chap25:sec:gwtc3}.
\textbf{Right panel}: Redshift distribution of the total number of observable BBHs, $N_\mathrm{exp}$, for ET.}
\label{chap25:fig:events_ET}
\end{figure}

The redshift-dependent observable events' number density of a GW detector can be calculated using
\begin{equation}
\label{chap25:eq:event_density} 
\frac{\mathrm{d} N_\mathrm{exp}}{\mathrm{d} z} = \int \mathrm{d} m_1 \mathrm{d} m_2\, \mathcal{R}(z, m_1, m_2)\,	\frac{\mathrm{d} VT}{\mathrm{d} z},
\end{equation} 
where $\frac{\mathrm{d} VT}{\mathrm{d} z}$ is defined in Eq.~\eqref{chap25:eq:dVTdz}.
Integrating over redshift $z$ yields the total number of observable events, $N_\mathrm{exp}$,
\begin{equation}
\label{chap25:eq:Nobs2} 
	N_\mathrm{exp} = \int \mathrm{d} z \frac{\mathrm{d} N_\mathrm{exp}}{\mathrm{d} z}.
\end{equation}
It's worth noting that Eq.~\eqref{chap25:eq:Nobs} is a special case of Eq.~\eqref{chap25:eq:Nobs2} when $P(m_1) = P(m_2) = \delta(m)$. Fig.~\ref{chap25:fig:events_ET} illustrates the expected number of observable BBHs $\mathrm{d} N_\mathrm{exp}/\mathrm{d} z$ and the total number of observable BBHs $N_\mathrm{exp}$, as a function of redshift for ET.

Third-generation GW detectors like ET are anticipated to detect $\mathcal{O}(10^5)$ BBH mergers annually and extend their reach to much higher redshifts. The distinct redshift distributions of $\mathrm{d} N_\mathrm{exp}/\mathrm{d} z$ for PBH binaries and ABH binaries provide a complementary tool to distinguish between these two BBH formation models. Particularly, for ABH models, the contribution to the expected number of observable BBHs from high redshifts ($z > 5$) may be negligible, leading to the total number of observable events, $N_\mathrm{exp}$, approaching a constant at $z > 5$. However, for PBH binaries, the contribution from higher redshifts cannot be ignored, causing $N_\mathrm{exp}$ to continue increasing when $z > 5$. Therefore, an ``excess" of the total number of observable events after redshift $z = 5$ could potentially indicate a population of PBHs as shown in Fig.~\ref{chap25:fig:events_ET}.

\section{Summary}
\label{chap25:sec:concs}

In summary, our exploration of the origins of BBH events detected by the LVK collaboration has led us to consider PBHs as potential progenitors, given their candidacy as dark matter constituents. Through a Bayesian analysis utilizing the LVK GWTC-3 catalog, we find that the stellar-mass PBHs are unlikely to constitute the entirety of CDM.

Our investigation extends to a mixed population of ABHs and PBHs, revealing that approximately one-fourth of the detectable events in GWTC-3 may be attributed to PBH binaries. This nuanced understanding of the BBH population sheds light on the coexistence of different astrophysical sources.

Looking ahead, we anticipate the third-generation GW detector, such as the ET, to provide valuable insights. By forecasting detectable event rate distributions for PBH and ABH binaries, this advanced detector offers a potential avenue to distinguish between PBHs and ABHs based on their distinct redshift evolutions. This promises to be a crucial step in refining our understanding of the various sources contributing to the rich tapestry of GW signals observed by GW detectors.

\begin{acknowledgement}
ZCC is supported by the National Natural Science Foundation of China (Grant No.~12247176 and No.~12247112) and the China Postdoctoral Science Foundation Fellowship No. 2022M710429. AH is supported by a Royal Society University Research Fellowship. For the purpose of open access, the author has applied a Creative Commons Attribution (CC BY) license to any Author Accepted Manuscript version arising from this submission.   
\end{acknowledgement}

\bibliographystyle{unsrt}
\bibliography{references.bib}

\end{document}